\documentclass[nojss]{jss}\usepackage[]{graphicx}\usepackage[]{xcolor}
\makeatletter
\def\maxwidth{ %
  \ifdim\Gin@nat@width>\linewidth
    \linewidth
  \else
    \Gin@nat@width
  \fi
}
\makeatother

\definecolor{fgcolor}{rgb}{0.345, 0.345, 0.345}
\newcommand{\hlnum}[1]{\textcolor[rgb]{0.686,0.059,0.569}{#1}}%
\newcommand{\hlsng}[1]{\textcolor[rgb]{0.192,0.494,0.8}{#1}}%
\newcommand{\hlcom}[1]{\textcolor[rgb]{0.678,0.584,0.686}{\textit{#1}}}%
\newcommand{\hlopt}[1]{\textcolor[rgb]{0,0,0}{#1}}%
\newcommand{\hldef}[1]{\textcolor[rgb]{0.345,0.345,0.345}{#1}}%
\newcommand{\hlkwb}[1]{\textcolor[rgb]{0.69,0.353,0.396}{#1}}%
\newcommand{\hlkwc}[1]{\textcolor[rgb]{0.333,0.667,0.333}{#1}}%
\newcommand{\hlkwd}[1]{\textcolor[rgb]{0.737,0.353,0.396}{\textbf{#1}}}%

\usepackage{framed}
\makeatletter
\newenvironment{kframe}{%
 \def\at@end@of@kframe{}%
 \ifinner\ifhmode%
  \def\at@end@of@kframe{\end{minipage}}%
  \begin{minipage}{\columnwidth}%
 \fi\fi%
 \def\FrameCommand##1{\hskip\@totalleftmargin \hskip-\fboxsep
 \colorbox{shadecolor}{##1}\hskip-\fboxsep
     \hskip-\linewidth \hskip-\@totalleftmargin \hskip\columnwidth}%
 \MakeFramed {\advance\hsize-\width
   \@totalleftmargin\z@ \linewidth\hsize
   \@setminipage}}%
 {\par\unskip\endMakeFramed%
 \at@end@of@kframe}
\makeatother

\definecolor{shadecolor}{rgb}{.97, .97, .97}
\definecolor{messagecolor}{rgb}{0, 0, 0}
\definecolor{warningcolor}{rgb}{1, 0, 1}
\definecolor{errorcolor}{rgb}{1, 0, 0}
\newenvironment{knitrout}{}{} 

\usepackage{alltt}
\usepackage[T1]{fontenc}
\usepackage[utf8]{inputenc}
\usepackage{lmodern}
\usepackage[american]{babel}  
\usepackage{bm,amsmath,amsfonts}

\newcommand{\bbeta}{{\bm \beta}}
\newcommand{\btheta}{{\bm \theta}}

\DeclareUnicodeCharacter{0394}{$\Delta$}
\shortcites{lesnoff_within-herd_2004}
\author{
  Anna Ly\\McMaster University \And
  Rune Haubo Bojesen Christensen\\Copenhagen Research Centre \\for Biological and Precision Psychiatry\AND
  Douglas Bates\\University of Wisconsin - Madison \And
  Martin M\"achler\\ETH Zurich\AND
  Benjamin M. Bolker\\McMaster University
}
\Plainauthor{A. Ly, R. H. B. Christensen, D. M. Bates, M. M\"achler, B. M. Bolker}
\title{Fitting generalized linear mixed-effects models using \pkg{lme4}}
\Plaintitle{Fitting generalized linear mixed models using lme4}
\Shorttitle{GLMMs with lme4}
\Abstract{%
  The \pkg{lme4} \proglang{R} package can be used to fit generalized linear
  mixed models (GLMMs), which extend the class of linear mixed models
  (LMMs). The two main extensions provided by GLMMs are (1)
  allowing for the conditional distribution of the response given the
  random effects to be non-Gaussian (e.g. binomial,
  Poisson) and (2) allowing the conditional mean to be
  a nonlinear function of a linear combination of the fixed
  and random effect coefficients, via an inverse link
  function. The conditional mode of the random effects given the
  observed data, the variance-covariance matrix of the random 
  effects, and the fixed effect parameters
  are determined using penalized iteratively reweighted
  least squares. We compute an approximation of the integral
  over the distributions of the conditional modes to compute
  the maximum likelihood estimate for a given set of parameters (by default we use the
  Laplace approximation or, alternatively, the more computationally
  expensive adaptive Gauss-Hermite quadrature). The package provides
  all the standard features available for GLMs in base \proglang{R},
  including the standard set of accessor functions as
  well as the possibility of user-specified distributions 
  (within the exponential dispersion family) and link
  functions.}
\Address{
  Anna Ly\\
  Department of Mathematics \& Statistics \\
  McMaster University \\
  1280 Main Street W \\
  Hamilton, ON L8S 4K1, Canada \\
  Email: \email{annahuynh.ly@utoronto.ca}
  \par\bigskip
  Rune Haubo Bojesen Christensen \\
  Copenhagen Research Centre for Biological and Precision Psychiatry \\
  Mental Health Centre Copenhagen, Copenhagen University Hospital - Bispebjerg and Frederiksberg \\
  Copenhagen, Denmark \\
  Email: \email{Rune.Haubo@pm.me}
  \par\bigskip
  Douglas Bates\\
  Department of Statistics, University of Wisconsin - Madison\\
  1205 University Ave. \\
  Madison, WI 53706, U.S.A.\\
  E-mail: \email{bates@stat.wisc.edu}
  \par\bigskip
  Martin M\"achler\\
  Seminar f\"ur Statistik, HG G~16\\
  ETH Zurich\\
  8092 Zurich, Switzerland\\
  E-mail: \email{maechler@stat.math.ethz.ch}\\
  \par\bigskip
  Benjamin M. Bolker\\
  Departments of Mathematics \& Statistics and Biology \\
  McMaster University \\
  1280 Main Street W \\
  Hamilton, ON L8S 4K1, Canada \\
  E-mail: \email{bolker@mcmaster.ca}
}

\newcommand{\mc}[1]{\ensuremath{\mathcal{#1}}}
\newcommand{\trans}{\ensuremath{^\top}}

\DeclareMathOperator*{\argmin}{arg\,min}

\usepackage{etoolbox}
\IfFileExists{upquote.sty}{\usepackage{upquote}}{}
\begin{document}
\section{Introduction}
\label{sec:intro}

The \pkg{lme4} package for \proglang{R} can be used to fit a broad
range of mixed-effects models.  One major advantage of
\pkg{lme4} over its predecessor, \pkg{nlme}, is that it can be used to
fit generalized linear mixed
models (GLMMs), which combine the flexibility of
linear mixed models (LMMs) and generalized
linear models (GLMs). In a companion paper, we have described the
facilities in \pkg{lme4} for fitting linear mixed models (LMMs). Here
we describe the facilities for fitting GLMMs.

\section{Generalized Linear Mixed Models}
\label{sec:GLMMdef}

Generalized linear mixed models extend the class of linear models in two ways: allowing 
for both fixed-effects parameters and random effects in the linear predictor, as in
linear mixed models (LMMs), and allowing for non-Gaussian distributions of the
response, whose mean is an inverse link applied to the linear predictor,
as in generalized linear models (GLMs).

We consider several characteristics of each of these model types.

\subsection{The Linear Model}
A linear model can be written as $\mc Y\sim\mc N\left(\bm X\bm\beta,\sigma^2\bm I\right)$, where
$\mc Y$ is the $n$-dimensional random variable representing the response, $\bm X$ is an $n\times p$
model matrix and $\bm\beta$ is a $p$-dimensional coefficient vector.
The vector $\bm\eta=\bm X\bm\beta$ (the \emph{linear predictor}) is mapped to
the \emph{mean vector}, $\bm\mu=\bm g^{-1}(\bm\eta)$, in this case using an \emph{identity} link and
inverse link function.
(The \emph{link} function, $\bm g$, maps $\bm\mu$ to $\bm\eta$ and the \emph{inverse link},
$\bm g^{-1}$, maps $\bm\eta$ to $\bm\mu$.)
The maximum likelihood estimates (MLEs) of the coefficients, $\widehat{\bm\beta}$, are the values that
minimize the (weighted) sum of squared residuals
\begin{equation}
  \label{eq:LMMmles}
  \widehat{\bm\beta} = \argmin_{\bm\beta}\sum_{i=1}^n w_i (y_i - \mu_i)^2 .
\end{equation}
with weights $w_i$.
The weighted squared residuals, $d(y_i,\mu_i, w_i)= w_i (y_i - \mu_i)^2$, are the \emph{unit deviances} for
the Gaussian family.\footnote{In the \proglang{R} \emph{family} lists of functions that encapsulate the
distribution families in arguments to \code{glm} and \code{glmer}, the function that returns the unit deviances
is named \code{dev.resids}, a source of confusion in some cases because they are not the values returned
by \code{residuals(m, type="deviance")}.}

\subsection{The Generalized Linear Model}
In a \emph{generalized linear model} (GLM) the linear predictor is still $\bm\eta=\bm X\bm\beta$
but the distribution of the response, $\mc Y$, with mean $\bm\mu=\bm g^{-1}(\bm\eta)$, can be from
distribution families other than the Gaussian, specifically from the class of exponential dispersion families; the most common examples are Bernoulli, binomial, or Poisson.
Individual components of the response, ${\mc Y}_i,i=1,\dots,n$, are pairwise independent.
The distribution family determines the \emph{unit deviance} function, $d(y_i,\mu_i)$, and the MLEs
of the coefficients are those that minimize the sum of the unit deviances
\begin{equation}
  \label{eq:sumofunitdeviances}
  \widehat{\bm\beta} = \argmin_{\bm\beta}\sum_{i=1}^n d(y_i,\mu_i, w_i) .
\end{equation}

The unit deviances are the (weighted) difference between negative twice the log-likelihood at $(y_i, \mu_i)$ and
negative twice the log-likelihood at $(y_i, y_i)$ (which is often zero).
For the Poisson distribution the unit deviance function is
\begin{equation}
  \label{eq:Poissonunitdeviance}
  d(y_i,\mu_i,w_i)=
  \begin{cases}
    2 w_i \mu_i & \text{if } y_i = 0 \\
    2 w_i (y_i \log(y_i/\mu_i) - (y_i - \mu_i)) & \text{otherwise}
  \end{cases}
\end{equation}
where $w_i$ is an optional prior case weight.
For simplicity we generally neglect the prior weights and write $d(y_i, u_i)$ below.

Minimizing the sum of the squared residuals in a linear model (i.e. finding the \emph{least squares} estimates
$\widehat{\bm\beta}$ for an LM) can be performed as a direct (i.e. non-iterative) calculation.
For a generalized linear model minimizing the sum of unit deviances usually must be done through iteration.
Fortunately, this can be done using a fast, robust algorithm, \emph{iteratively reweighted least squares} \cite[IRLS:][]{McCullaghNelder1989}.

The link and inverse link functions are \emph{diagonal mappings}, in the sense that there is a scalar
function, $g$, such that the $i$th component of $\bm\eta$ is $g$
applied to the $i$th component of $\bm\mu_{\mc Y}$.  (The name
``diagonal'' reflects the fact that the Jacobian matrix,
$\frac{d\eta}{d\mu\trans}$, of such a mapping will be diagonal.)
If the distribution family is one of the exponential distribution families a
\emph{canonical link} can be derived from the form of the distribution.
The canonical link is the identity for the Gaussian family, the \emph{logit} or ``log-odds'' function,
$g(\mu)=\log(\mu/(1 - \mu))$, for the Bernoulli or binomial families, and the
natural logarithm, $g(\mu)=\log(\mu)$, for the Poisson.

\subsection{The Linear Mixed Model}
In a linear mixed model (LMM), described in \cite{bates2015fitting}, there are two vector-valued random
variables: $\mc Y$, the $n$-dimensional response, and $\mc B$, the $q$-dimensional random effects vector.
The unconditional distribution $\mc B\sim\mc N(\bm 0,\bm\Sigma_{\bm\theta})$ depends on a
\emph{covariance parameter} $\bm\theta$.
The conditional distribution, $(\mc Y| \mc B=\bm b)\sim\mc N(\bm X\bm\beta+\bm Z\bm b, \sigma^2\bm I)$,
has mean, $\bm\mu$, from the identity link applied to the linear predictor $\bm\eta=\bm X\bm\beta+\bm Z\bm b$,
where $\bm Z$ is an $n\times q$ model matrix for the random effects and $\bm b$ is a value of $\mc B$.
In practice the covariance parameter, $\bm\theta$, determines a lower triangular \emph{relative covariance factor},
$\bm\Lambda_{\bm\theta}$, with $\bm\Sigma_{\bm\theta}=\sigma^2\bm\Lambda_{\bm\theta}\bm\Lambda_{\bm\theta}\trans$
and we work with a ``spherical'' random effects variable, $\mc U\sim\mc N(\bm 0,\sigma^2\bm I)$, such that
$\mc B=\bm\Lambda_{\bm\theta}\mc U$, and for which the linear predictor is
\begin{equation}
  \label{eq:LMMlinpred}
  \bm\eta=\bm X\bm\beta+\bm Z\bm b=\bm X\bm\beta+\bm Z\bm\Lambda_{\bm\theta}\bm u .
\end{equation}
($\mc U$ is said to be ``spherical'' because contours of constant probability density
are spheres centered at the origin.)

Most generally, $\bm\Lambda_{\bm\theta}$ can be any lower triangular matrix, since
$\bm\Lambda_{\bm\theta}\bm\Lambda_{\bm\theta}\trans$ is then guaranteed to be
positive semidefinite. In \pkg{lme4} version 2.0 and later, we have also
implemented \emph{structured covariance matrices}, which introduce a more
elaborate mapping from $\bm\theta$ to $\bm\Lambda_{\bm\theta}$ that lets users
specify a range of structured forms such as diagonal, compound symmetric, or
first-order autoregressive (AR1) covariance matrices. We introduce the syntax
for these structures in the CBPP example (Section~\ref{sec:cbpp}); see the
\href{https://CRAN.R-project.org/package=lme4/vignettes/covariance_structures.html}{lme4 covariance structures vignette}
for the details of how $\bm\Lambda_{\bm\theta}$ is constructed for
each structure.

The likelihood of the parameters, $\bm\beta$, $\bm\theta$ and $\sigma^2$, given $\mc Y=\bm y$,
is the marginal density of $\mc Y$ evaluated at the observed $\bm y$.
Conceptually, we determine an expression for the integral of the joint density,
$f_{\mc Y,\mc U}(\bm y,\bm u)$, with respect to $\bm u$, then evaluate that expression at the observed $\bm y$.
In practice, it is easier to reverse the order of the evaluation at the observed $\bm y$ and the integration with
respect to $\bm u$.

The joint density is
\begin{equation}
  \label{eq:jointLMMdensity}
  f_{\mc Y,\mc U}(\bm y,\bm u)=\frac{1}{(2\pi\sigma^2)^{(n+q)/2}}
  \exp\left(\frac{\|\bm y-\bm\mu\|^2 + \|\bm u\|^2}{-2\sigma^2}\right) ,
\end{equation}
where the expression in the numerator of the integrand is the \emph{penalized sum of the unit deviances} (PSUD)
for the Gaussian family, which is the penalized sum of squared residuals (PSSR).
(The sum of the unit deviances is $\|\bm y-\bm\mu\|^2$ and the penalty is $\|\bm u\|^2$.)

To evaluate the integral we determine the \emph{conditional mode} of $\mc U$, given $\mc Y=\bm y$, which minimizes
the PSSR with respect to $\bm u$
\begin{equation}
  \label{eq:penalizedsumofsquares}
  (\tilde{\bm u} | \mc Y=\bm y,\bm\theta,\bm\beta) = \argmin_{\bm u}\left[\sum_{i=1}^n(y_i-\mu_i)^2+\|\bm u\|^2\right] .
\end{equation}
The PSSR can be minimized directly (i.e. without iteration) and the minimum doesn't depend on $\sigma^2$.
When the minimum PSSR and the determinant of the Hessian of the PSSR, which can be evaluated from a Cholesky
factor created in the process of solving eq.~\ref{eq:penalizedsumofsquares}, are available the MLE of
$\sigma^2$ can be evaluated separately, as is commonly done in linear models.

Because $\bm\beta$ and $\bm u$ both occur as coefficients in the linear predictor, the conditional estimate,
$\widehat{\bm\beta}|\bm\theta$, can be determined along with $\tilde{\bm u}$
by extending the penalized least squares problem,
(eq.~\ref{eq:penalizedsumofsquares}), to minimize with respect to both $\bm u$ and $\bm\beta$.
Thus, given a value of $\bm\theta$ and the solution to this extended PSSR problem,
the profiled log-likelihood, which is a function of $\bm\theta$ only, can be evaluated.
General nonlinear optimizers applied to the profiled log-likelihood as a function of $\bm\theta$ are
used to determine the MLEs for all the parameters.

The point of using the profiled log-likelihood instead of the original log-likelihood is that the
dimension of the general nonlinear optimization problem is much smaller in the profiled problem.

\subsection{The Generalized Linear Mixed Model}

As stated earlier, a GLMM combines the distribution family and link function, from a GLM, with random
effects in the linear predictor, as in an LMM.
The unconditional distribution of the random effects, $\mc B\sim\mc N(\bm 0, \bm\Sigma_{\bm\theta})$,
and the definitions of the spherical random effects, $\mc U$, such that $\mc B=\bm\Lambda_{\bm\theta}\mc U$ where
$\Lambda_{\bm\theta}\Lambda_{\bm\theta}\trans=\bm\Sigma$ are as in the LMM except that the scale parameter,
$\sigma$, is not present in the most common GLMMs (binomial and Poisson families).
The mean of the conditional distribution, $\mc Y|\mc U=\bm u$, is the inverse link applied to the
linear predictor $\bm\eta=\bm X\bm\beta+\bm Z\bm\Lambda_{\bm\theta}\bm u$.

The conditional mode, which minimizes the \emph{penalized sum of unit deviances} (PSUD),
\begin{equation}
  \label{eq:condmodeGLMM}
  (\tilde{\bm u}|\mc Y=\bm y,\bm\theta,\bm\beta)=\argmin_{\bm u}\left[\sum_{i=1}^n d(y_i,\mu_i) + \|\bm u\|^2\right],
\end{equation}
is determined by penalized iteratively reweighted least squares (PIRLS).

In the case of an LMM, minimizing the PSSR provided the value of the integral defining the likelihood of
the parameters, given the observed response, $\mc Y=\bm y$. 
For a GLMM, minimizing the PSUD provides a local quadratic approximation to the logarithm of the joint density,
$f_{\mc Y,\mc U}(\bm y,\bm u)$.
Approximating the integral with respect to $\bm u$ of the joint density (evaluated at the observed $\bm y$)
by the integral of this local quadratic approximation is \emph{Laplace's method}.

For simple models involving scalar random effects from a single random-effects term the multivariate integral
with respect to $\bm u$ can be factored into a product of scalar integrals, which can be evaluated by
Gauss-Hermite rules.


To summarize:
\begin{enumerate}
\item Like a GLM, the definition of a GLMM incorporates a distribution family and a link function.
  The distribution family determines the unit deviances, $d(y_i,\mu_i),i=1,\dots,n$. Distributions in the exponential distribution family have a canonical link which is the default link in \code{glm} and in \code{glmer} for those families.
\item As in an LMM the linear predictor, $\bm\eta=\bm X\bm\beta+\bm Z\bm b$, incorporates fixed-effects      
  parameters, $\bm\beta$, and random effects, $\mc B\sim\mc N(\bm 0,\bm\Sigma_{\bm\theta})$.
\item The parameters $\bm\theta$ determine a triangular relative covariance factor, $\bm\Lambda_{\bm\theta}$,
  such that $\bm\Sigma_{\bm\theta}=\bm\Lambda_{\bm\theta}\bm\Lambda_{\bm\theta}\trans$.
  (A scale parameter, $\sigma$, is incorporated for families like the Gaussian that use one.)
\item For given values of $\bm\beta$ and $\bm\theta$, the penalized sum of unit deviances (PSUD),
  $\sum_{i=1}^n d(y_i,\mu_i)+\|\bm u\|^2$, is minimized with respect to $\bm u$ using PIRLS.
\item A local quadratic approximation to the PSUD at the conditional mode provides Laplace's approximation
  to the log-likelihood, which is minimized with respect to $\bm\beta$ and $\bm\theta$.
\item In certain simple models adaptive Gauss-Hermite quadrature (aGHQ) can be used to
  approximate the log-likelihood more accurately, at the expense of more evaluations of the PSUD for each
  candidate combination of $\bm\beta$ and $\bm\theta$.
\item If the PSUD is minimized with respect to $\bm u$ and $\bm\beta$ simultaneously then the approximate
  profiled log-likelihood becomes a function of $\bm\theta$ only, as discussed below.
  This approximation can save computation time for large problems.
\end{enumerate}

\subsection{Determining the conditional mode}
\label{sec:conditionalMode}

The iteratively reweighted least squares (IRLS) algorithm is an
efficient method of determining the maximum likelihood
estimates of the coefficients in a GLM.
As the name IRLS implies this is an iterative algorithm where the $(k+1)$st
coefficient vector, $\bm\beta^{(k+1)}$, is determined from $\bm\beta^{(k)}$, and the process
continues until convergence.
At the $k$th iteration, the linear predictor is $\bm\eta^{(k)}=\bm X\bm\beta^{(k)}$ and
the mean response vector is $\bm\mu^{(k)}=\bm g^{-1}(\bm\eta^{(k)})$.
The residual on the response scale, $\bm r^{(k)}=\bm y-\bm\mu^{(k)}$, is mapped to a \emph{working residual}
on the linear predictor scale via a linear approximation to the link function.
Because the link and inverse link are diagonal maps we can write the working residual component-wise as
\begin{equation}
  \label{eq:wrkresid}
  \tilde{r}_i^{(k)}=(y_i - \mu_i^{(k)})g^\prime(\mu_i^{(k)})\quad i=1,\dots,n ,
\end{equation}
where $g^\prime(\mu_i^{(k)})$ denotes the derivative of the link function at $\mu_i^{(k)}$.

Because the variances of the working residuals are not constant we must evaluate these variances
and use weighted least squares, with the weights inversely proportional to the variances, to determine an
increment $\bm\delta^{(k)}$ such that $\bm\beta^{(k+1)}=\bm\beta^{(k)}+\bm\delta^{(k)}$. (The \emph{working weights} based on the variances are combined with the prior weights $w_i$, if any have been specified.)
An alternative approach is to define a \emph{working response}
\begin{equation}
  \label{eq:wrkresp}
  \tilde{y}_i^{(k)}=(\eta_i^{(k)}- o_i)+\tilde{r}_i^{(k)},\quad i=1,\dots,k
\end{equation}
where $o_i$ is an optional \emph{offset} to be applied on the linear predictor scale.

IRLS is usually implemented by solving a weighted least squares problem on the working residual
for the increment, $\bm\delta^{(k)}$, at each iteration.
For the \emph{penalized iteratively reweighted least squares} (PIRLS) algorithm to determine
$\bm u^{(k+1)}$ from $\bm u^{(k)}$ it is easiest to use the working response because the penalty
is defined as $\|\bm u^{(k+1)}\|^2$ so the penalized weighted least squares calculation should be
in terms of $\bm u$.

In more detail, the PIRLS algorithm has the form
\begin{enumerate}
\item Given parameter values, $\bm\beta$ and $\bm\theta$, incorporate the fixed-effects contribution
  to the linear predictor, $\bm X\bm\beta$, as an \emph{offset}, $\bm o$, to be used in evaluating the working
  response (eq. \ref{eq:wrkresp}).
  Set the initial values of $\bm u^{(0)}=\bm 0$ so that the PIRLS iterations always start from the same value
  of $\bm u$, providing for a stable result that depends only on $\bm\beta$ and $\bm\theta$.
  Evaluate the lower triangular $\bm\Lambda_{\bm\theta}$, the linear predictor
  \begin{equation}
    \label{eq:linpredPIRLS0}
    \bm\eta^{(0)}=\bm Z\bm\Lambda_{\bm\theta}\bm u^{(0)} + \bm o,
  \end{equation}
  the conditional mean of $\mc Y|\mc U=\bm u^{(0)}$,
  \begin{equation}
    \label{eq:condmeanPIRLS0}
    \bm\mu^{(0)}=\bm g^{-1}(\bm\eta^{(0)}),
  \end{equation}
  the (diagonal) Jacobian matrix,
  \begin{equation}
    \label{eq:Jacobian}
    \bm J^{(0)}=\frac{d \bm\mu}{d \bm\eta^\top}={\bm g^{-1}}^\prime
  \end{equation}
  and the conditional variance
  \begin{equation}
    \label{eq:confvarPIRLS}
    \bm V^{(0)}=\mathrm{Var}(\mc Y|\bm\mu=\bm\mu^{(0)})
  \end{equation}
  which is also a diagonal matrix.
  
\item Establish the weights as the inverse of the combined variance contributions from $\bm\eta$:
  \begin{equation}
    \label{eq:PIRLSweights}
    \bm W^{(0)}=\left(\bm J^{(0)\top}\bm V^{(0)}\bm J^{(0)}\right)^{-1}
  \end{equation}
(we would also multiply this quantity by a diagonal matrix of prior weights, if any were specified).
  Eq. \ref{eq:PIRLSweights} is written as a product and an inverse of matrices but all the matrices involved
  are diagonal and the evaluation of eq.~\ref{eq:PIRLSweights} is performed as vector operations.

\item Evaluate $\bm u^{(1)}$ as the solution to the penalized, weighted least squares problem
  \begin{equation}
    \label{eq:PIRLSeqn}
    \left(\bm \Lambda^\top\bm Z^\top\bm W^{(0)}\bm Z\bm\Lambda + \bm I\right)\bm u^{(1)}=
    \bm\Lambda^\top\bm Z^\top \bm W^{(0)}\tilde{\bm y}^{(0)}
  \end{equation} 
  using the sparse Cholesky factor, $\bm L_{\bm\beta,\bm\theta}^{(0)}$, defined by
  \begin{equation}
    \label{eq:sparseChol}
    \bm L_{\bm\beta,\bm\theta}^{(0)}\bm L_{\bm\beta,\bm\theta}^{(0)\top}=
    \bm P\left(\bm \Lambda^\top\bm Z^\top\bm W^{(0)}\bm Z\bm\Lambda + \bm I\right)\bm P^\top ,
  \end{equation}
  where $\bm P$ is the matrix representation of a \emph{fill-reducing permutation} determined from the
  structure of $\bm Z^\top\bm Z$.
  (In simple GLMM problems $\bm P$ can be an identity matrix but for general sparse Cholesky evaluations
  determining and employing a fill-reducing permutation can be important in reducing memory usage
  and compute time.)
  
\item Evaluate PSUD, the penalized sum of unit deviances, at $\bm u^{(1)}$.
  If the PSUD at $\bm u^{(1)}$ exceeds that at $\bm u^{(0)}$ use \emph{step-halving} which, in this case,
  means replacing $\bm u^{(1)}$ by the average of $\bm u^{(0)}$ and $\bm u^{(1)}$, and trying again.
  We set a maximum, usually 10 or fewer, on the number of step-halvings that are allowed before
  declaring the algorithm to have failed.

\item If the iteration succeeds in reducing the PSUD, check for convergence by comparing the weights,
  $\bm W^{(i+1)}$, to $\bm W^{(i)}$.
  If the relative differences in the weights are small the algorithm is declared to have converged.
  Otherwise repeat the previous steps using evaluations at $\bm u^{(i)}$ to produce $\bm u^{(i+1)}$.

\end{enumerate}

\subsection{Evaluating the likelihood for GLMMs using the Laplace approximation}
\label{sec:Laplace}

Evaluating the likelihood for generalized linear mixed models
requires approximating an intractable 
integral over the random effects distribution. The \code{glmer} function offers 
several approximations, controlled by the \code{nAGQ} argument. 
The default value of \code{nAGQ=1} specifies the
\emph{Laplace approximation} \citep{madsen2011}.

A second-order Taylor series approximation to $-2\log[f_{\mc Y,\mc
  U}(\bm y,\bm u)]$ based at $\tilde{\bm u}$ provides an approximation
of the unscaled conditional density as a multiple of the density for
the multivariate Gaussian $\mathcal{N}(\tilde{\bm u},\bm L\bm
L\trans)$.  The change of variable
\begin{equation}
  \label{eq:LaplaceChg}
  \bm u = \tilde{\bm u} + \bm L\bm z
\end{equation}
provides
\begin{equation}
  \label{eq:GLMMLaplace}
  \begin{aligned}
    L(\bm\beta,\bm\theta|\bm y)&=\int_{\mathbb{R}^q}f_{\mc Y,\mc U}(\bm y,\bm u)\,d\bm u\\
    &\approx \tilde{f}\,|\bm L|\, \int_{\mathbb{R}^q}e^{-\|\bm z\|^2/2}\,(2\pi)^{-q/2}\,d\bm z\\
    &=\tilde{f}\, |\bm L|
  \end{aligned}
\end{equation}
(where $\tilde f$ is the penalized deviance evaluated at the conditional modes)
or, on the deviance scale,
\begin{equation}
  \label{eq:LaplaceDev}
  -2\ell(\bm\beta,\bm\theta|\bm y)\approx\sum_{i=1}^n d(y_i, \mu_i) +
    \|\tilde{\bm u}\|^2 + \log(|\bm L|^2)+\frac{q}{2}\log(2\pi)
\end{equation}

The Laplace approximation normally conditions on both the fixed effects $\beta$ and the variance-covariance parameters $\theta$. A further approximation, which is denoted in \code{glmer} by \code{nAGQ=0}, profiles out the fixed effects by minimizing with respect to $\bm\beta$ and $\bm u$ simultaneously in the PIRLS algorithm (i.e., replacing $\bm u$ with $(\bm \beta \; \bm u)$ and $\bm Z \bm \Lambda$ with $(\bm X | \bm Z \bm \Lambda)$ throughout, except with the penalty term still applying only to $\bm u$). This approximation is exact when (1) $\partial(\log L)/\partial \beta$ is a linear function of the conditional modes $\bm u$ and (2) when the conditional mode is equal to the conditional mean (typically, although not necessarily, implying a symmetric conditional distribution). Both assumptions hold for linear mixed models (although a Laplace approximation is not necessary there), consistent with \cite{bates2015fitting} showing that the fixed effects can be profiled out of the log-likelihood for LMMs. The Julia \code{MixedModels.jl} package offers the same approximation as the \code{fast} argument to the \code{pirls!} function (\url{https://juliastats.org/MixedModels.jl/stable/optimization/}); Template Model Builder \citep{kristensenTMB2016}, and downstream packages such as \code{glmmTMB}, provide this function via a \code{profile} argument.

By default, \code{glmer} uses a two-stage optimization procedure (described below) with \code{nAGQ=0} in the first stage; users can also specify \code{nAGQ=0} for faster, approximate model fits.

\subsubsection{Decomposing the deviance for simple models}
\label{sec:simplescalar}

A common special case of mixed models is those where scalar (typically
intercept) random effects are
associated with levels of a single grouping factor, $\bm h$.  In this
case the dimension, $q$, of the random effects is the number of levels
of $\bm h$ --- i.e.{} there is exactly one random effect associated
with each level of $\bm h$.  We will write the vector of
variance-covariance parameters, which is one-dimensional, as a scalar,
$\theta$.  The matrix $\bm\Lambda_{\bm\theta}$ is a multiple of the
identity, $\theta\bm I_q$, and $\bm Z$ is the $n\times q$ matrix of
indicators of the levels of $\bm f$.  The permutation matrix, $\bm
P$, can be set to the identity and $\bm L$ is diagonal, although not
necessarily homogeneous (i.e., a scalar multiple of the identity matrix).

Because each element of $\bm\mu$ depends on only one element of $\bm
u$ and the elements of $\mc Y$ are conditionally independent, given
$\mc U=\bm u$, the conditional densities of the $u_j,j=1,\dots,q$
given $\mc Y=\bm y$ are independent.  We partition the indices
$1,\dots,n$ as $\mathbb{I}_j,j=1,\dots,q$ according to the levels of
$\bm h$.  That is, the index $i$ is in $\mathbb{I}_j$ if $h_i=j$.
This partitioning also applies to the deviance residuals in that
the $i$th deviance residual depends only on $u_j$ when $i\in\mathbb{I}_j$.

Writing the univariate conditional densities as
\begin{equation}
  \label{eq:univariateCondDens}
  f_j(\bm y,u_j)=\exp\left(-\frac{\sum_{i\in\mathbb{I}_j}d(y_i, u_j)+u_j^2}{2}\right)(2\pi)^{-1/2}
\end{equation}
we have
\begin{equation}
  \label{eq:vectorCondDens}
  f_{\mc Y,\mc U}(\bm y,\bm u)=\prod_{j=1}^q f_j(\bm y,u_j)
\end{equation}
and
\begin{equation}
  \label{eq:ssLike}
  \begin{aligned}
    L(\bm\beta,\bm\theta|\bm y)=\prod_{j=1}^q\int_{\mathbb{R}}f_j(\bm y,u)\,du
  \end{aligned}
\end{equation}

We consider this special case both because it occurs frequently and
because, for some software, it is the only type of GLMM that can be
fit.  Also, in this particular case we can graphically assess the
quality of the Laplace approximation by comparing the actual integrand
to its approximation.

Consider the \code{cbpp} data on contagious bovine pleuropneumonia
(CBPP)
incidence according to season and herd, available in the \pkg{lme4} package
(see \ref{sec:cbpp} for more details), and the model
\begin{knitrout}
\definecolor{shadecolor}{rgb}{0.969, 0.969, 0.969}\color{fgcolor}\begin{kframe}
\begin{alltt}
\hldef{> }\hlkwd{print}\hldef{(m1} \hlkwb{<-} \hlkwd{glmer}\hldef{(}\hlkwd{cbind}\hldef{(incidence, size}\hlopt{-}\hldef{incidence)} \hlopt{~} \hldef{period} \hlopt{+} \hldef{(}\hlnum{1}\hlopt{|}\hldef{herd),}
\hldef{+ }      \hldef{cbpp, binomial),} \hlkwc{corr}\hldef{=}\hlnum{FALSE}\hldef{)}
\end{alltt}
\begin{verbatim}
Generalized linear mixed model fit by maximum likelihood (Laplace
  Approximation) [glmerMod]
 Family: binomial  ( logit )
Formula: cbind(incidence, size - incidence) ~ period + (1 | herd)
   Data: cbpp
      AIC       BIC    logLik -2*log(L)  df.resid 
 194.0531  204.1799  -92.0266  184.0531        51 
Random effects:
 Groups Name        Std.Dev.
 herd   (Intercept) 0.6421  
Number of obs: 56, groups:  herd, 15
Fixed Effects:
(Intercept)      period2      period3      period4  
    -1.3983      -0.9919      -1.1282      -1.5797  
\end{verbatim}
\end{kframe}
\end{knitrout}
This model has been fit by minimizing the Laplace approximation to the
deviance.  We can assess the quality of this approximation by
evaluating the unscaled conditional density at $u_j(z)=\tilde{u_j} +
z/{\bm L_{j,j}}$ and comparing the ratio,
$f_j(\bm y,u)/(\tilde{f_j}\sqrt{2\pi})$, to the standard normal
density, $\phi(z)=e^{-z^2/2}/\sqrt{2\pi}$, as shown in Figure~\ref{fig:densities}.

\begin{figure}[tbp]
  \centering
\begin{knitrout}
\definecolor{shadecolor}{rgb}{0.969, 0.969, 0.969}\color{fgcolor}
\includegraphics[width=\maxwidth]{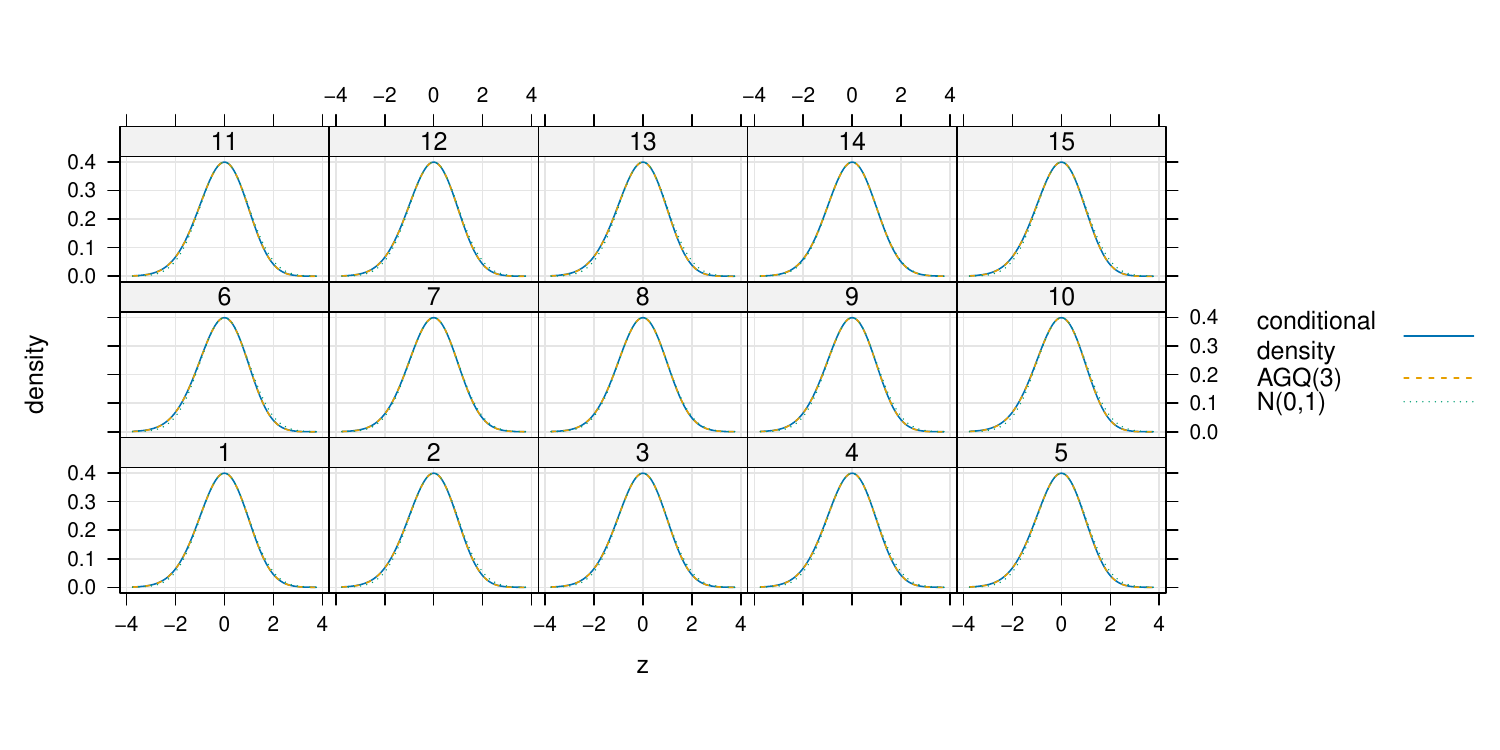} 
\end{knitrout}
\caption{Comparison of univariate integrands (solid, blue line); 3-point Gauss-Hermite quadrature (dashed, yellow); and standard normal density function (green, dotted) for the CBPP model. For this model, the three approximations are nearly indistinguishable.}
  \label{fig:densities}
\end{figure}
As Figure~\ref{fig:densities} shows, the univariate integrands are very
close to the standard normal density, indicating that the Laplace
approximation to the deviance is a good approximation in this case.

\section{Adaptive Gauss-Hermite quadrature for GLMMs}
\label{sec:aGQ}
When the first integral in (\ref{eq:GLMMLaplace}) can be expressed as a product of
low-dimensional integrals, we can use Gauss-Hermite quadrature to
provide a closer approximation to the integral.  Univariate
Gauss-Hermite quadrature evaluates the integral of a function that is
multiplied by a ``kernel'' where the kernel is a multiple of
$e^{-z^2}$ or $e^{-z^2/2}$.  For statisticians the natural candidate
is the standard normal density, $\phi(z)=e^{-z^2/2}/\sqrt(2\pi)$.
A $k$th-order Gauss-Hermite formula provides knots, $z_i,i=1,...,k$,
and weights, $w_i,i=1,\dots,k$, such that
\begin{equation}
  \label{eq:GHquad}
  \int_{\mathbb{R}}t(z)\phi(z)\,dz\approx\sum_{i=1}^kw_it(z_i)
\end{equation}
The function \code{GHrule} in \pkg{lme4} (based on code in the
\pkg{SparseGrid} package) provides knots and weights relative to the
standard normal kernel for orders $k$ from 1 to 100.  For example,
\begin{knitrout}
\definecolor{shadecolor}{rgb}{0.969, 0.969, 0.969}\color{fgcolor}\begin{kframe}
\begin{alltt}
\hldef{> }\hlkwd{GHrule}\hldef{(}\hlnum{5}\hldef{)}
\end{alltt}
\begin{verbatim}
             z          w     ldnorm
[1,] -2.856970 0.01125741 -5.0000774
[2,] -1.355626 0.22207592 -1.8377997
[3,]  0.000000 0.53333333 -0.9189385
[4,]  1.355626 0.22207592 -1.8377997
[5,]  2.856970 0.01125741 -5.0000774
\end{verbatim}
\end{kframe}
\end{knitrout}
where \code{z} is the vector of knots, \code{w} is the vector
of weights, and \code{ldnorm} is the log-density of the standard
normal distribution at $z$.

The choice of the value of $k$ depends on the behavior of the function
$t(z)$.  If $t(z)$ is a polynomial of degree $2k-1$ then the
Gauss-Hermite formula for orders $k$ or greater provides an exact
answer.  The fact that we want $t(z)$ to behave like a low-order
polynomial is often neglected in the formulation of a Gauss-Hermite
approximation to a quadrature.  The quadrature knots on the $u$ scale
are chosen as
\begin{equation}
  \label{eq:quadraturepts}
  u_{i,j}(z)=\tilde{u_j} + z_i/{\bm L_{j,j}},\quad i=1,\dots,k;\;j=1,\dots,q
\end{equation}
exactly so that the function $t(z)$ should behave like a low-order
polynomial over the region of interest, which is to say the region
where quadrature knots with large weights are located.  The term
``adaptive Gauss-Hermite quadrature'' reflects the fact that the
approximating Gaussian density is scaled and shifted to provide a
second order approximation to the logarithm of the unscaled
conditional density.

\begin{figure}[tbp]
  \centering
\begin{knitrout}
\definecolor{shadecolor}{rgb}{0.969, 0.969, 0.969}\color{fgcolor}
\includegraphics[width=\maxwidth]{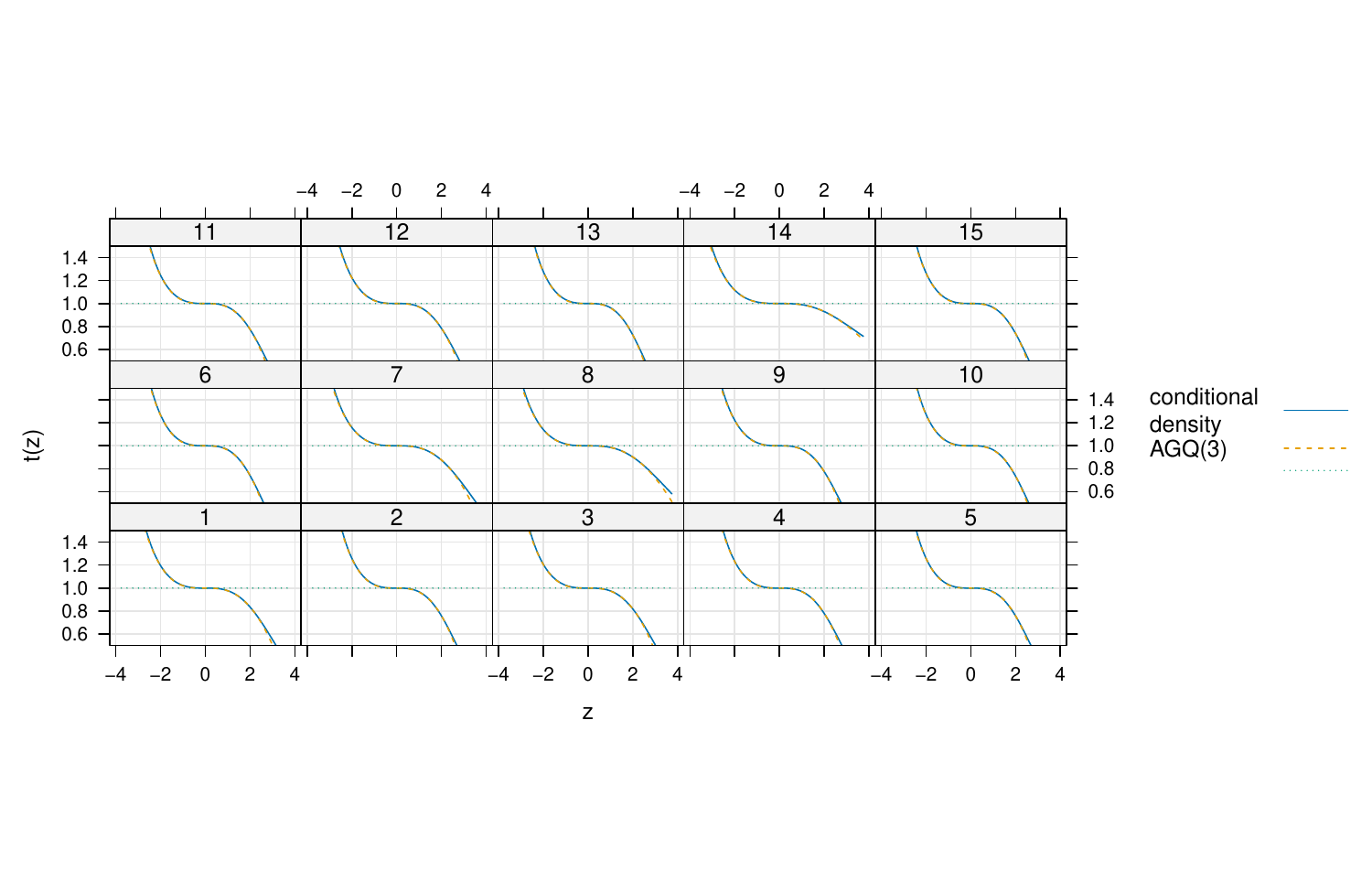} 
\end{knitrout}
  \caption{The function $t(z)$, which is the ratio of the normalized
    unscaled conditional density to the standard normal density, for
    each of the univariate integrals in the evaluation of the deviance
    for model \code{m1} (blue, solid line).
    As in Figure~\ref{fig:densities}, the dashed yellow line shows
    the approximation for 3-point adaptive Gauss-Hermite quadrature.
    The horizontal green reference line ($z=1$)
    gives the reference value that would apply for the Laplace approximation,
    which assumes the conditional density is exactly equal to the
    standard normal. The $y$-axis limits are truncated to (0.5, 1.5).
}
  \label{fig:tfunc}
\end{figure}

Figure~\ref{fig:tfunc}
shows $t(z)$ for each of the unidimensional integrals in the
likelihood for the model \code{m1} at the parameter estimates.
While this view shows the deficiency of the Laplace approximation
(deviation from a horizontal reference line at $z=1$) clearly,
The AGQ(3) approximation fits extremely well over the range shown.
The tails of the polynomials implied by the AGQ approximation
can fluctuate widely, but these fluctuations are suppressed by
multiplying by the thin tails of the Gaussian distribution (so that
in Figure~\ref{fig:densities} the deviations of the conditional density
from the standard Normal are barely visible).

\begin{figure}[tbp]
  \centering
\begin{knitrout}
\definecolor{shadecolor}{rgb}{0.969, 0.969, 0.969}\color{fgcolor}
\includegraphics[width=\maxwidth]{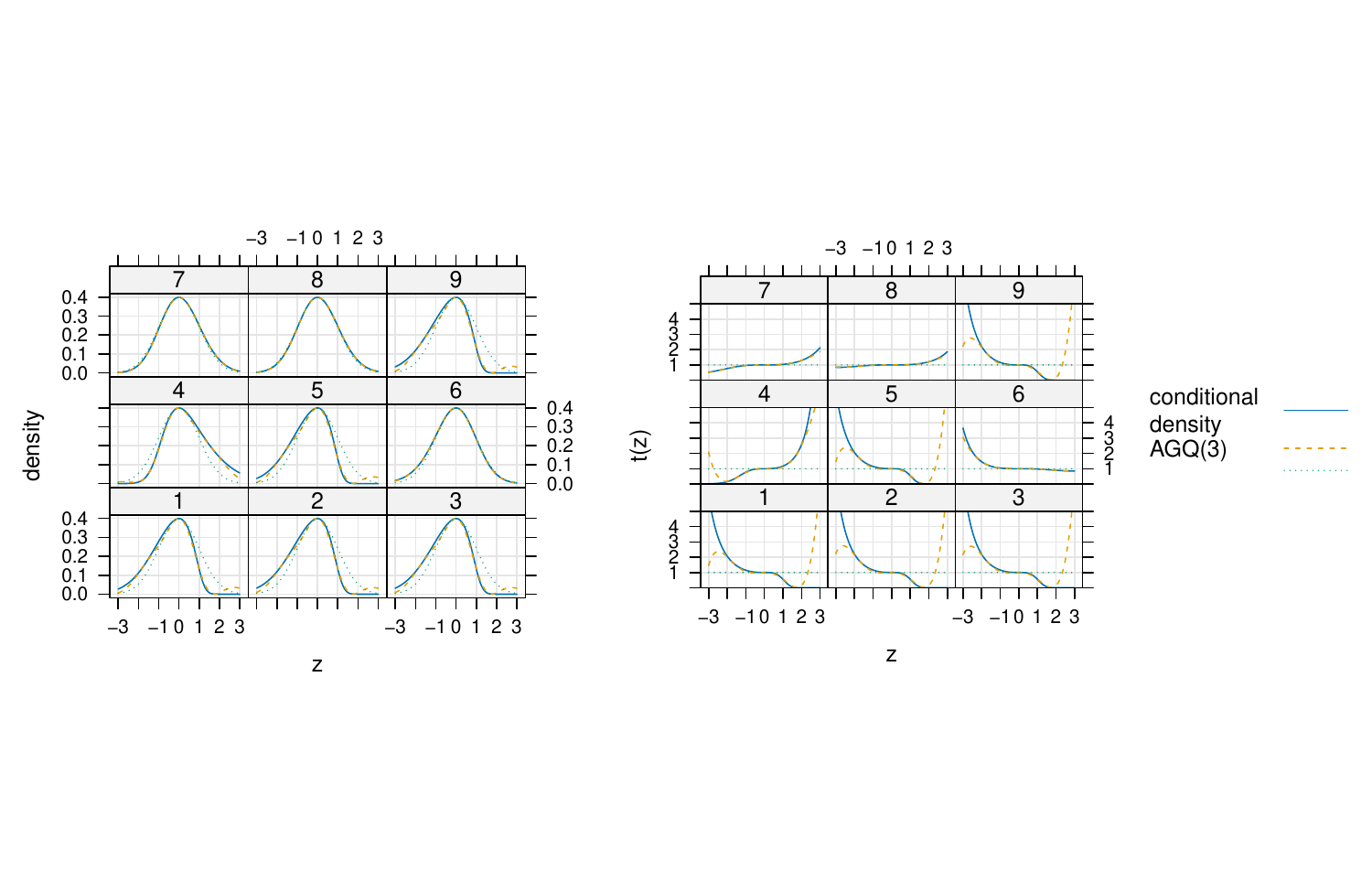} 
\end{knitrout}
\caption{Normalized unscaled conditional density (left) and ratio of density to the standard normal density (right)
  for a random sample of 9 patients from the toenail onychomycosis data set. The $y$-axis limits in the right panel ($t(z)$) are
  truncated to $(0,5)$.}
\label{fig:toenailplot}
\end{figure}

The CBPP data set is a relatively well-behaved data set, where Laplace approximation works well. In contrast,
a widely used data set on toenail onychomycosis \citep{debacker+1998}, which has a very low effective sample size per cluster --- an
average of about 6.5 binary observations (``moderate or severe'' vs ``none or mild'' disease) per patient --- represents
an example where Gauss-Hermite quadrature is necessary for reliable results. In this case the
conditional densities depart much more clearly from the standard normal (Figure~\ref{fig:toenailplot}; note the scale of the
density ratios goes from 0 to 10, in contrast the maximum of 4 in Figure~\ref{fig:tfunc}).
\cite{stringerAsymptoticsNumericalIntegration2022} and \cite{stringerExactGradientEvaluation2024} further explore the limitations of Laplace approximation.

To use adaptive Gauss-Hermite quadrature for model fitting in \code{glmer} models, 
users would set the argument \code{nAGQ}, the number of quadrature points 
(i.e. $k$ in eq.~\ref{eq:GHquad}), to a value greater than 1.

Increasing the number of nodes generally improves the accuracy of the likelihood 
approximation at the expense of computation time --- although rounding errors
may accumulate when using large numbers of quadrature points.
A reasonable strategy for choosing the number of nodes is to re-fit the model with increasing values of \code{nAGQ} until the parameter estimates stabilize
\citep{tuerlinckxStatisticalInferenceGeneralized2006a}; a ``warm-start'' procedure using the parameter estimates from the previous fit as starting values (and using \code{control = glmerControl(nAGQ0initStep = FALSE)} to disable the \code{nAGQ=0} preliminary fit) will speed up this procedure.

At present, AGQ is only available 
for models with a single scalar random effect. (The \code{GLMMadaptive} package implements
AGQ for vector-valued random effect models, although it is still restricted to models with
a single random effect; \cite{rabe-heskethMaximumLikelihoodEstimation2005b} describe methods
for AGQ for models with nested random effects.)

\section{Model fitting}

  Once we can calculate the deviance by PIRLS for specified values of $\btheta$ and $\bbeta$ (or only $\btheta$ if profiling out the fixed effects via \code{nAGQ=0}), we then estimate the parameters by nonlinear optimization. This procedure largely follows the description in \cite{bates2015fitting}, using derivative-free optimizers with box constraints to prevent non-positive-(semi)definite covariance matrices. Specifically, the elements of $\btheta$ corresponding to the diagonal of $\bm\Lambda_\theta$ are currently constrained to be non-negative.
  
The only difference is that by default \code{glmer} uses a two-step fitting procedure, using \code{nAGQ=0} at the first stage to get preliminary estimates which are then used as starting points for a second optimization with Laplace approximation or Gauss-Hermite quadrature as specified by the user. Different nonlinear optimizers can be used at each stage: the current default, as specified in \code{glmerControl}, is to use Powell's BOBYQA \citep{Powell_bobyqa} followed by a box-constrained variant of the Nelder-Mead simplex algorithm. For faster, approximate fitting, the second stage can be omitted; in the rare cases where the initial \code{nAGQ=0} fit gives poor results, the first stage can be skipped via \code{glmerControl(nAGQ0initStep = FALSE)}.
  
However, because the profiled log-likelihood is an even function of the diagonal elements of $\btheta$
and $\bm\Sigma_\theta=\bm\Lambda_\theta\bm\Lambda_\theta\trans$ depends on them only through their squares, positive and negative values yield identical likelihoods. This symmetry means that constrained optimization is not strictly necessary --- an unconstrained optimizer will converge to a correct solution, approaching zero from either side in the boundary case of a singular random-effects covariance matrix. (Once a solution is found, we can map it to a unique solution where the diagonal elements are all non-negative.)
A similar approach to removing constraints could work for structured covariance matrices where correlation parameters are constrained to $(-1,1)$, e.g. by parameterizing the model in terms of a phase parameter $p$ where $\rho = \sin(p)$ (and then mapping $p$ to $(0, 2\pi)$).
Removing these constraint would permit the use of a broader class of unconstrained optimizers, though this feature has not yet been incorporated into \code{lme4}.

\section{Examples}

\subsection{CBPP}
\label{sec:cbpp}

The \code{?cbpp} help page describes the CBPP data set
\citep{lesnoff_within-herd_2004} as follows:
\begin{quote}
Contagious bovine pleuropneumonia (CBPP) is a major disease of
cattle in Africa, caused by a mycoplasma.  This dataset describes
the serological incidence of CBPP in zebu cattle during a
follow-up survey implemented in 15 commercial herds located in the
Boji district of Ethiopia.  The goal of the survey was to study
the within-herd spread of CBPP in newly infected herds. Blood
samples were quarterly collected from all animals of these herds
to determine their CBPP status.  These data were used to compute
the serological incidence of CBPP (new cases occurring during a
given time period).  Some data are missing (lost to follow-up).
\end{quote}
\cite{lesnoff_within-herd_2004} estimated
the effects of different treatments using (1) ordinary logistic regression
incorporating a variance-inflation factor, also known as
a quasi-binomial model (``logistic regression'' is sometimes
used specifically to describe analyses of Bernoulli responses, but
in this case there are multiple trials per observation [cows that
could become seropositive], and so a dispersion or scale parameter
can be estimated); (2) a GLMM implemented in \code{lme4}; and a
(3) Markov chain Monte Carlo algorithm \citep{zeger1991generalized}, 
which as they state allows for a non-parametric rather than a Normal model 
for the random effects.
The authors did not find any significant effects of treatment,
ascribing the null results to 
``a lack of power in the statistical analyses or to a quality problem for the medications used (and more generally, for health-care delivery in the Boji district).''

(Note that Table 1 of \cite{lesnoff_within-herd_2004} contains a known
typographical error for herd 6. Consequently, results obtained using
the \code{cbpp} data set may not exactly reproduce some of the findings
reported in that paper.)

The \code{lme4} package includes two variants of the \code{cbpp}
data set. The second variant, \code{cbpp2}, contains corrected values
corresponding to Table~1 of \cite{lesnoff_within-herd_2004}. Although
the precise provenance of these data sets is unclear, the \code{cbpp}
data set matches the version held by the corresponding author of
\cite{lesnoff_within-herd_2004}.

\begin{knitrout}
\definecolor{shadecolor}{rgb}{0.969, 0.969, 0.969}\color{fgcolor}\begin{figure}
\includegraphics[width=\maxwidth]{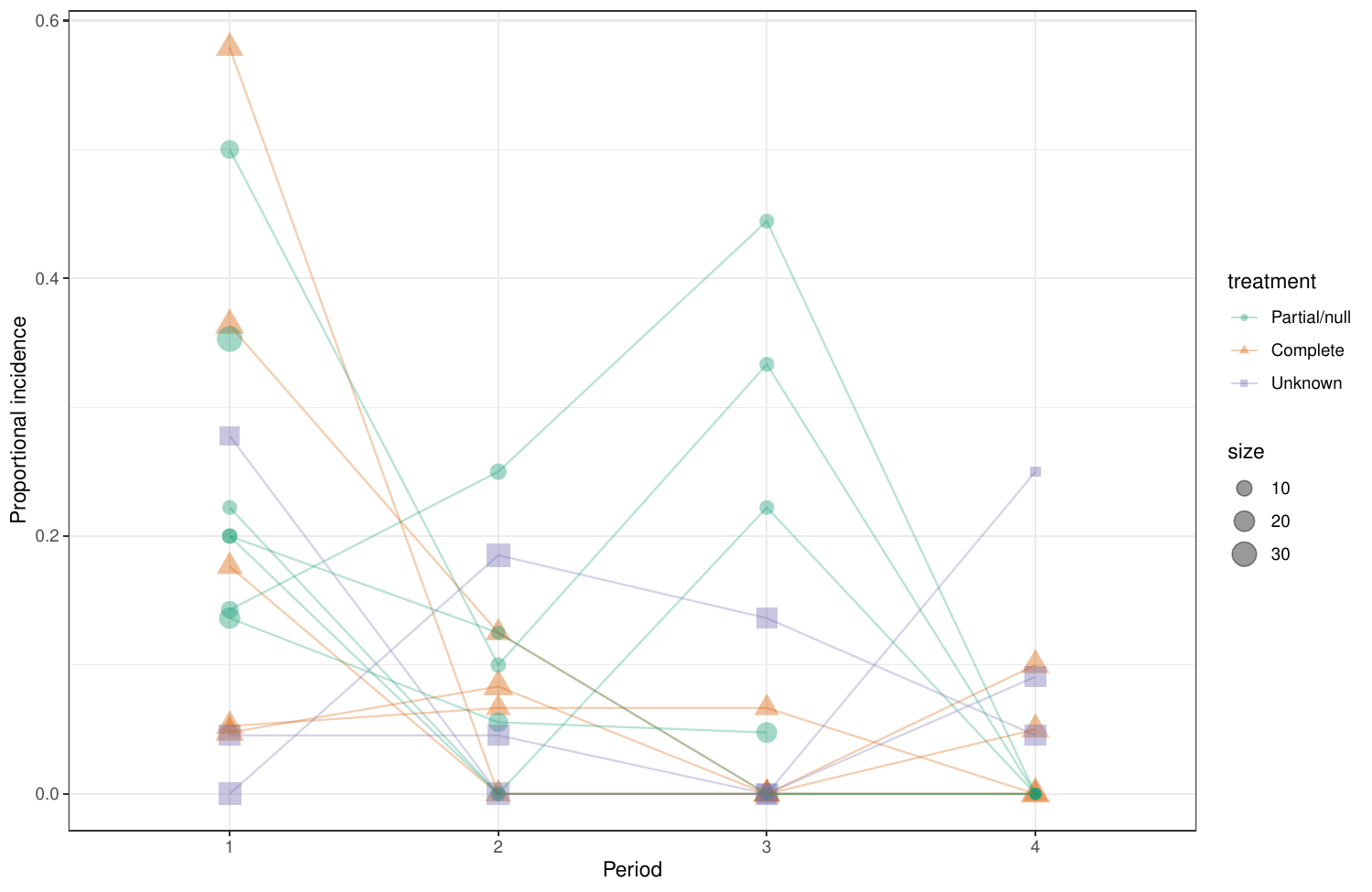} \caption[Incidence]{Incidence (proportion of cows becoming seropositive per observation period) vs. period. Colours show treatment category for each herd; point sizes reflect the number of seronegative cows at the start of each period. Lines connect the sets of observations from each herd.}\label{fig:cbppPlot}
\end{figure}

\end{knitrout}

We model proportional incidence as a binomial response depending
on the additive fixed effects of period, treatment and average herd size;
to account for repeated measures we fit a model with a random effect
of herd. In practice we might also be interested in a period by treatment
interaction, but we neglect that term here.

As in \code{glm}, we can specify 
a binomial response as a proportion and use the \code{weights}
argument to specify the sample size, instead of the
more typical two-column \code{cbind(successes,failures)} format:
\begin{knitrout}
\definecolor{shadecolor}{rgb}{0.969, 0.969, 0.969}\color{fgcolor}\begin{kframe}
\begin{alltt}
\hldef{> }\hldef{gm1} \hlkwb{<-} \hlkwd{glmer}\hldef{(incidence}\hlopt{/}\hldef{size} \hlopt{~} \hldef{period} \hlopt{+} \hldef{treatment} \hlopt{+} \hldef{avg_size} \hlopt{+} \hldef{(}\hlnum{1} \hlopt{|} \hldef{herd),}
\hldef{+ }             \hlkwc{family} \hldef{= binomial,}
\hldef{+ }             \hlkwc{data} \hldef{= cbpp2,} \hlkwc{weights} \hldef{= size)}
\end{alltt}
\end{kframe}
\end{knitrout}

It is also worth considering
adding an observation-level random effect to the model \citep{harrisonComparisonObservationlevelRandom2015},
which we can do by creating a new factor based on observation
number and using \code{update()} on the previous model (we switch the optimizer to BOBYQA for both phases
for one of the updates):
\begin{knitrout}
\definecolor{shadecolor}{rgb}{0.969, 0.969, 0.969}\color{fgcolor}\begin{kframe}
\begin{alltt}
\hldef{> }\hldef{cbpp2} \hlkwb{<-} \hlkwd{transform}\hldef{(cbpp2,}\hlkwc{obs}\hldef{=}\hlkwd{factor}\hldef{(}\hlkwd{seq}\hldef{(}\hlkwd{nrow}\hldef{(cbpp2))))}
\hldef{> }\hldef{bob_opt} \hlkwb{<-} \hlkwd{glmerControl}\hldef{(}\hlkwc{optimizer} \hldef{=} \hlsng{"bobyqa"}\hldef{)}
\hldef{> }\hlcom{## herd and observation-level REs}
\hldef{> }\hldef{gm2} \hlkwb{<-} \hlkwd{update}\hldef{(gm1,.}\hlopt{~}\hldef{.}\hlopt{+}\hldef{(}\hlnum{1}\hlopt{|}\hldef{obs),} \hlkwc{control} \hldef{= bob_opt)}
\hldef{> }\hlcom{## observation-level REs only}
\hldef{> }\hldef{gm3} \hlkwb{<-} \hlkwd{update}\hldef{(gm1,.}\hlopt{~}\hldef{.}\hlopt{-}\hldef{(}\hlnum{1}\hlopt{|}\hldef{herd)}\hlopt{+}\hldef{(}\hlnum{1}\hlopt{|}\hldef{obs))}
\end{alltt}
\end{kframe}
\end{knitrout}

\subsubsection{Model summary}

The first part of the summary reiterates
the family and link function used,
the model formula, and gives various summary statistics
(log-likelihood etc.), as well as quantiles of the
scaled (Pearson) residuals:
\begin{knitrout}
\definecolor{shadecolor}{rgb}{0.969, 0.969, 0.969}\color{fgcolor}\begin{kframe}
\begin{verbatim}
Generalized linear mixed model fit by maximum likelihood (Laplace
  Approximation) [glmerMod]
 Family: binomial  ( logit )
Formula: incidence/size ~ period + treatment + avg_size + (1 | herd)
   Data: cbpp2
Weights: size

      AIC       BIC    logLik -2*log(L)  df.resid 
    197.8     214.0     -90.9     181.8        48 

Scaled residuals: 
    Min      1Q  Median      3Q     Max 
-2.2311 -0.7967 -0.3732  0.4684  2.7557 
\end{verbatim}
\end{kframe}
\end{knitrout}
These quantities are also
accessible via standard accessors
(\code{AIC()}, \code{BIC()}, \code{logLik()}).

The next chunk of \code{summary()} describes the random effects
and the number of levels associated with each grouping factor
(the latter is useful for checking that random-effects formulae
have been specified correctly):
\begin{knitrout}
\definecolor{shadecolor}{rgb}{0.969, 0.969, 0.969}\color{fgcolor}\begin{kframe}
\begin{verbatim}
Random effects:
 Groups Name        Variance Std.Dev.
 herd   (Intercept) 0.3116   0.5582  
Number of obs: 56, groups:  herd, 15
\end{verbatim}
\end{kframe}
\end{knitrout}
This information is also accessible via \code{VarCorr()}, which returns
a list of variance-covariance matrices (the \code{print} method
for \code{VarCorr} objects allows control of whether the variance,
or standard deviation, or both, are printed).

Next come the estimates of the fixed effects, along with Wald
estimates of the standard error, $Z$ statistic, and $p$-value:
\begin{knitrout}
\definecolor{shadecolor}{rgb}{0.969, 0.969, 0.969}\color{fgcolor}\begin{kframe}
\begin{verbatim}
Fixed effects:
                   Estimate Std. Error z value Pr(>|z|)
(Intercept)       -1.005623   0.708418  -1.420 0.155743
period2           -0.986283   0.303381  -3.251 0.001150
period3           -1.125147   0.323142  -3.482 0.000498
period4           -1.561098   0.422631  -3.694 0.000221
treatmentComplete -0.376225   0.503000  -0.748 0.454483
treatmentUnknown  -0.683246   0.645179  -1.059 0.289599
avg_size          -0.006135   0.045608  -0.135 0.893002
\end{verbatim}
\end{kframe}
\end{knitrout}
One can use \code{coef(summary())} to retrieve this information,
and optionally format it with \code{printCoefmat()}.

The last component of \code{summary()} gives the
estimated correlations among the fixed-effect parameters,
which can be useful for assessing multicollinearity
(it can also be overwhelming: it is suppressed by default
for models with more than 20 fixed-effect parameters, and
can also be suppressed by using \code{print(summary(.),correlation=FALSE)}).
\begin{knitrout}
\definecolor{shadecolor}{rgb}{0.969, 0.969, 0.969}\color{fgcolor}\begin{kframe}
\begin{verbatim}
Correlation of Fixed Effects:
            (Intr) perid2 perid3 perid4 trtmnC trtmnU
period2     -0.135                                   
period3     -0.130  0.278                            
period4     -0.086  0.210  0.195                     
trtmntCmplt  0.289 -0.017 -0.021 -0.059              
trtmntUnknw  0.431 -0.053 -0.045 -0.043  0.588       
avg_size    -0.910  0.026  0.028  0.020 -0.547 -0.649
\end{verbatim}
\end{kframe}
\end{knitrout}

\subsubsection{Diagnostics}

A range of graphical diagnostic tools is available
for \code{merMod} objects. The plot methods in the \code{lme4}
package are inspired by those in the \code{nlme} package, using
\code{lattice} plots to provide a reasonable blend
of convenience and flexibility. 

\code{merMod} objects are also compatible with the 
\code{performance} package and the \code{DHARMa} package, both commonly 
used for model checking.

The following code produces a standard range
of diagnostic plots (Figure~\ref{fig:glmerDiag}), similar to the ones
in base R's \code{plot.lm} method. These diagnostics will generally be
useful for models where the conditional density is approximately normal
(but heteroscedastic) --- e.g., Poisson responses with large mean or
binomial responses with large numbers of successes --- and less so otherwise,
e.g. for binary responses.

\begin{knitrout}
\definecolor{shadecolor}{rgb}{0.969, 0.969, 0.969}\color{fgcolor}\begin{kframe}
\begin{alltt}
\hldef{> }\hlcom{## basic residual plot}
\hldef{> }\hlkwd{plot}\hldef{(gm1)}
\hldef{> }\hlcom{## scale-location plot}
\hldef{> }\hlkwd{plot}\hldef{(gm1,}\hlkwd{sqrt}\hldef{(}\hlkwd{abs}\hldef{(}\hlkwd{resid}\hldef{(.)))}\hlopt{~}\hlkwd{fitted}\hldef{(.),}\hlkwc{type}\hldef{=}\hlkwd{c}\hldef{(}\hlsng{"p"}\hldef{,}\hlsng{"smooth"}\hldef{))}
\hldef{> }\hlcom{## boxplot of residuals grouped by a categorical predictor}
\hldef{> }\hlkwd{plot}\hldef{(gm1,period}\hlopt{~}\hlkwd{resid}\hldef{(.))}
\hldef{> }\hlcom{## Q-Q plot}
\hldef{> }\hlkwd{qqmath}\hldef{(gm1)}
\end{alltt}
\end{kframe}
\end{knitrout}

\begin{knitrout}
\definecolor{shadecolor}{rgb}{0.969, 0.969, 0.969}\color{fgcolor}\begin{figure}
\includegraphics[width=\maxwidth]{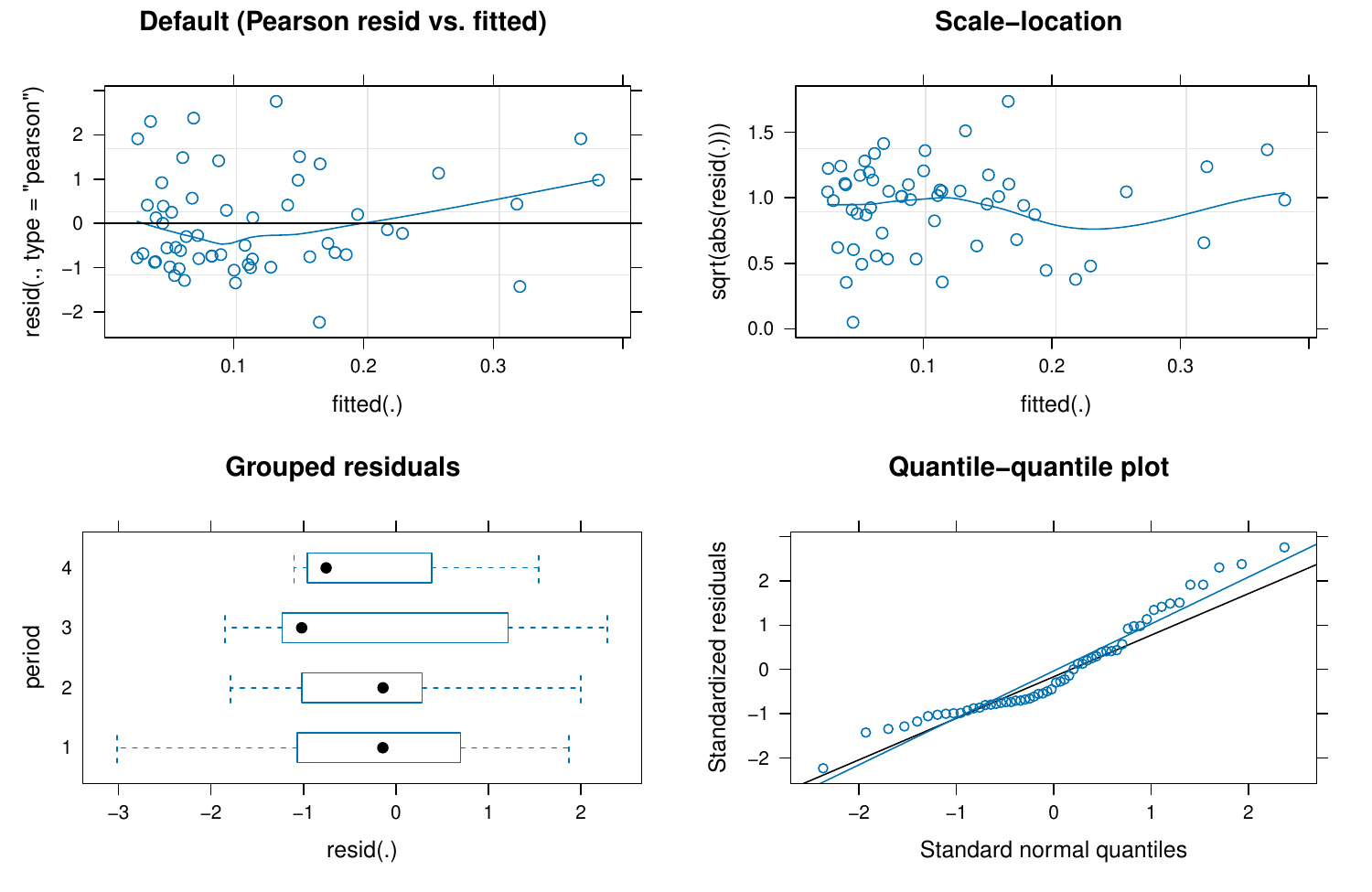} \caption[Graphical diagnostics (using package built-in functions)]{Graphical diagnostics (using package built-in functions)}\label{fig:glmerDiag}
\end{figure}

\end{knitrout}

The \code{ranef()} accessor extracts the conditional modes; the argument
\code{condVar=TRUE} additionally extracts the variances of
the conditional modes, which are stored as an attribute
labelled \code{"postVar"}\footnote{In earlier versions of the software, these elements were called ``posterior variances'' rather than ``conditional variances'', based on the close correspondence between mixed models and hierarchical Bayesian models.} --- a three-dimensional array that
gives the variance-covariance matrix of the conditional modes
for each level of the grouping variable. The plotting methods
\code{dotplot()} and \code{qqmath()} return lists of graphical
objects showing \emph{caterpillar plots} (ordered values of
the random effects with confidence bars); in the case of
the Q-Q plot (\code{qqmath}) the $y$-axis shows corresponding 
values of the standard normal quantiles (Figure~\ref{fig:glmerRanefplot}).

\begin{knitrout}
\definecolor{shadecolor}{rgb}{0.969, 0.969, 0.969}\color{fgcolor}\begin{figure}
\includegraphics[width=\maxwidth]{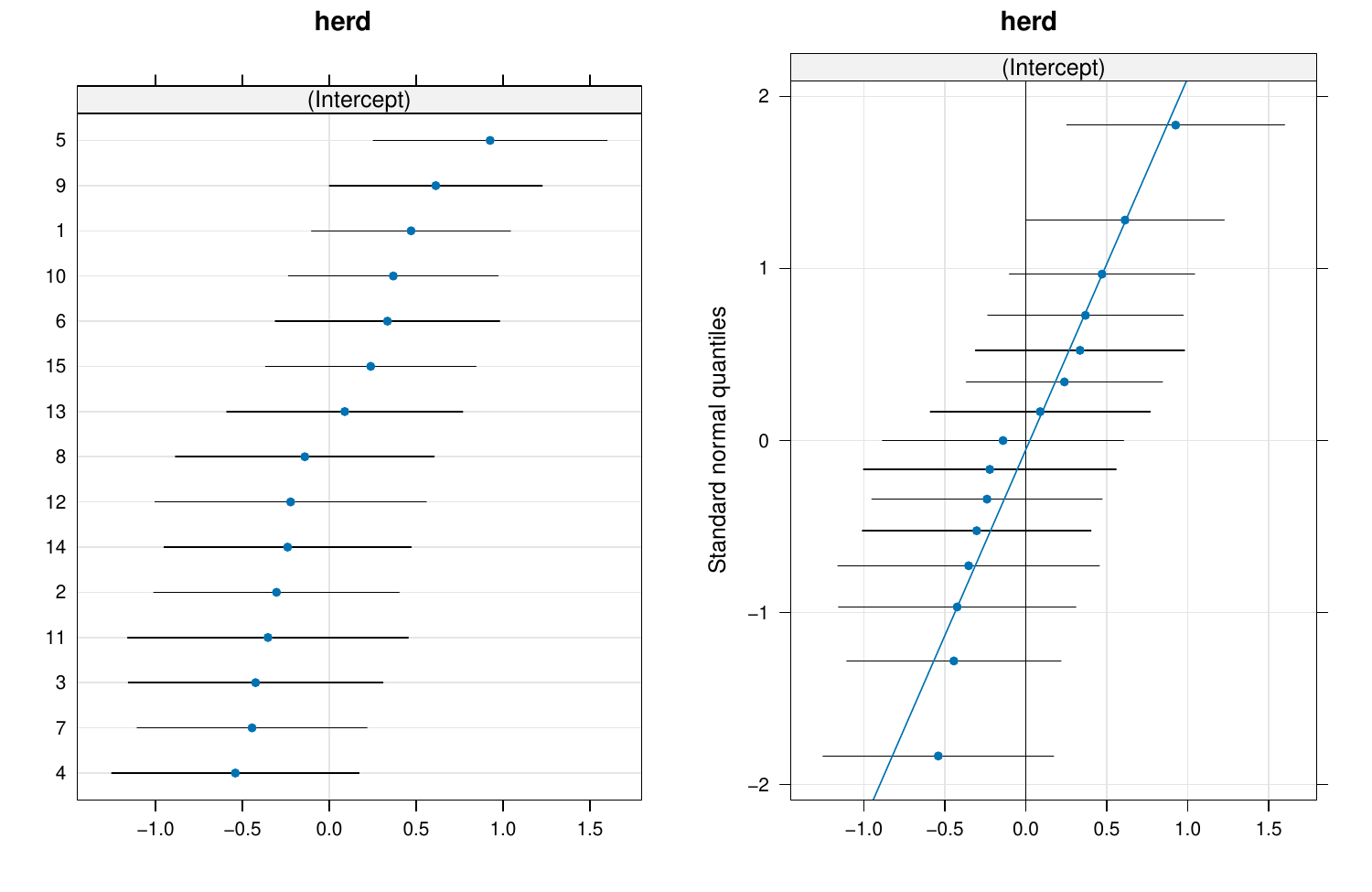} \caption{Graphical display of random effects. \emph{Left}: conditional modes $\pm 1.96 \times \text{conditional standard deviation}$, ordered by magnitude. \emph{Right}: quantile-quantile plot, with linear regression line overlaid.}\label{fig:glmerRanefplot}
\end{figure}

\end{knitrout}

\code{performance::check\_model()} checks a variety of classical assumptions 
of GLMs. For mixed-effect models, it also assesses
the normality of the distribution of conditional modes (Figure~\ref{fig:perf_chk_mod_cbpp}).

\begin{knitrout}
\definecolor{shadecolor}{rgb}{0.969, 0.969, 0.969}\color{fgcolor}\begin{figure}
\includegraphics[width=\maxwidth]{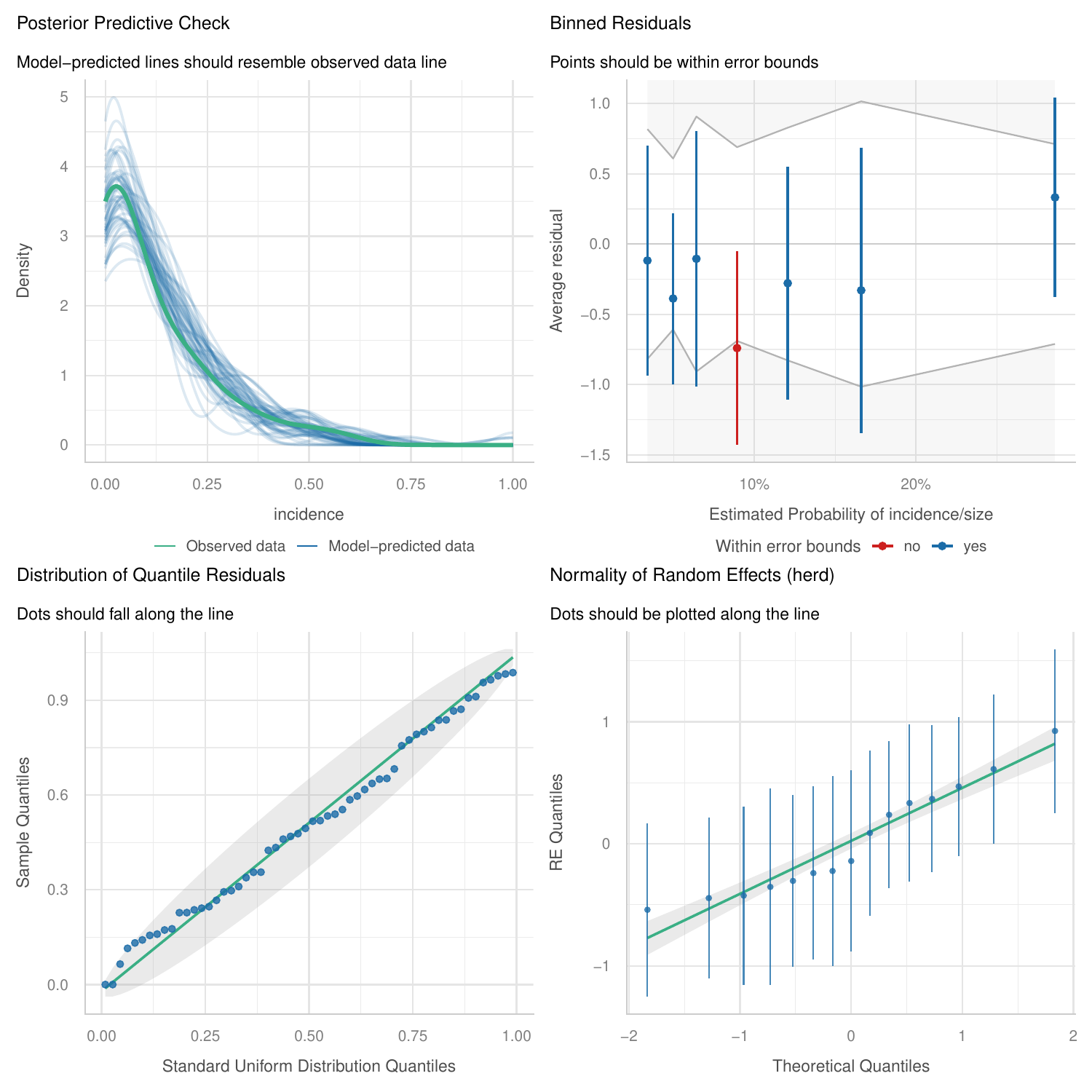} \caption{Model diagnostics using \code{performance::check\_model}. The plot showcasing the normality of random effects (lower right panel) is the transpose of the right panel in Figure 6.}\label{fig:perf_chk_mod_cbpp}
\end{figure}

\end{knitrout}

The \code{DHARMa} package is also compatible with \code{merMod} objects.
\code{DHARMa} generates simulation-based residuals for
generalized linear (mixed) models and uses them for graphical
and statistical tests of model assumptions (Figure~\ref{fig:dharma_mod_cbpp}).

\begin{knitrout}
\definecolor{shadecolor}{rgb}{0.969, 0.969, 0.969}\color{fgcolor}\begin{figure}
\includegraphics[width=\maxwidth]{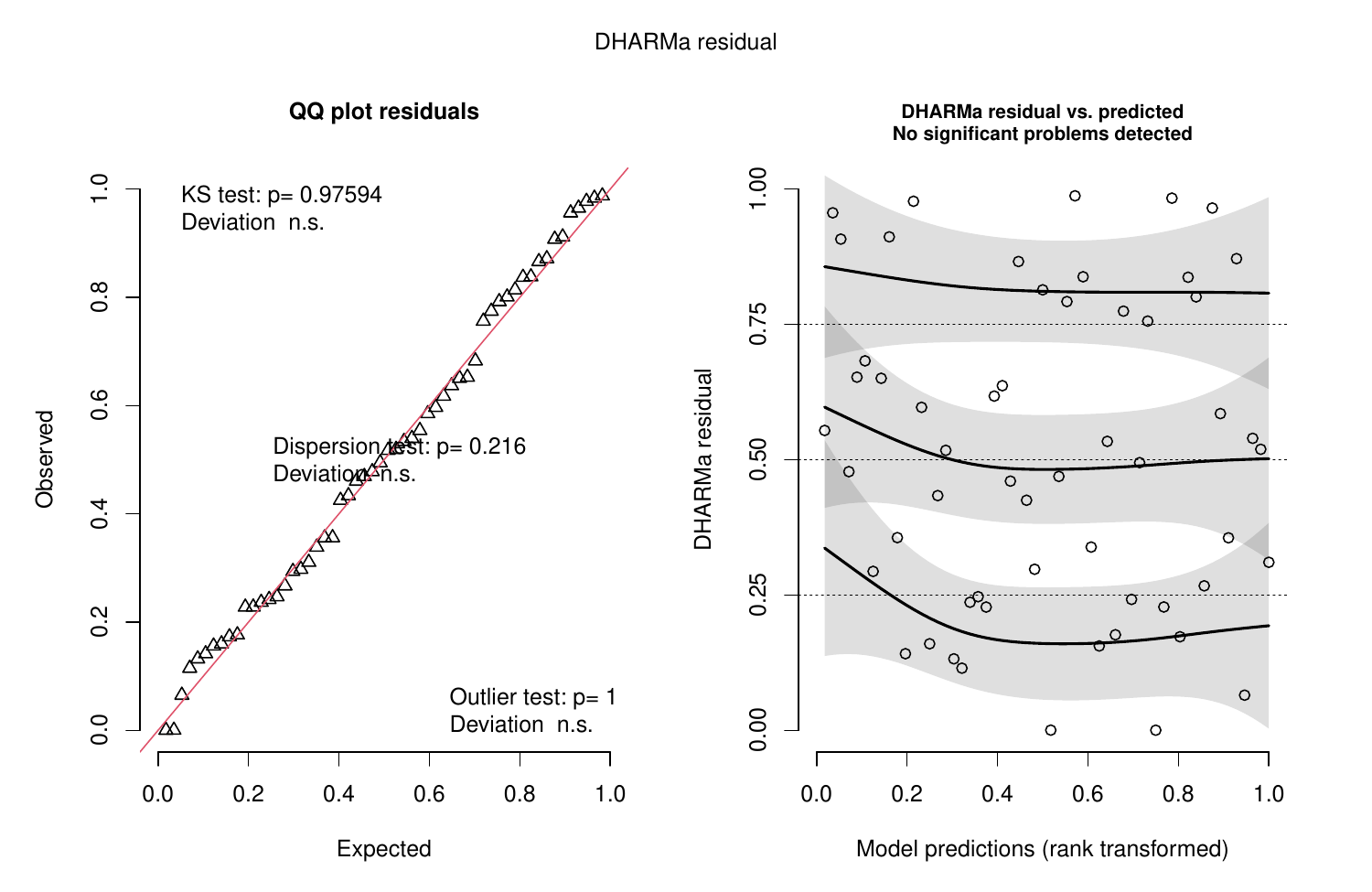} \caption{Model assumption checks using the DHARMa package. \code{n.s.} denotes ``not significant''.}\label{fig:dharma_mod_cbpp}
\end{figure}

\end{knitrout}

Having checked the diagnostics, we would now like to compare
the three models we have fitted.  Inspecting the \code{VarCorr}
components, we see that when we fit both herd- and observation-level
random effects, the among-herd variance is estimated as zero.
The appropriate procedure
at this point (e.g. whether one
drops non-significant terms, or those with small scaled magnitudes,
or those that worsen the AIC or BIC of the model)
depends on the goals of the analysis and one's
philosophy of model-building \citep{barrRandomEffectsStructure2013,matuschekBalancingTypeError2017,scandola2024}.

One might either stick with the full model, or continue
with the reduced model with observation-level random effects
only (as it has exactly the same likelihood as the full
model but uses an additional parameter, it would be chosen
according to either an information-theoretic or a
hypothesis-testing model selection framework).

Here we will start by
computing likelihood profiles and confidence intervals (CIs)
for the model incorporating both random effects;
although it has the same point estimates and maximum likelihood
as the reduced model, confidence intervals that incorporate
non-local information (i.e. profile- or parametric bootstrap-based)
will give different, more conservative results for the full model.


The \code{profile} method computes profile likelihoods.
The computation can be slow, since complete profiling for
a model with $p$ random- and fixed-effect parameters requires fitting $p$
profiles, each of which requires many $p-1$-dimensional optimizations.  The machinery for generating likelihood profiles for GLMMs is similar to that for LMMs (see \cite{bates2015fitting}, \S\ 5.1).

The \code{profile} method returns an
object of class \code{thpr} --- a data frame
containing the profiles, augmented with attributes containing
interpolation splines for each parameter profile and their inverses
(using \code{splines::interpSpline} and \code{splines::backSpline});
the latter are used for plotting profiles and computing confidence
intervals.  An \code{as.data.frame} method adds \code{.focal} and
\code{.par} variables to the data frame, useful for customized
plots.

Profiles can be used for univariate (\code{xyplot}) and
bivariate (\code{splom}) profile plots, and to compute
profile confidence intervals (\code{confint}). (\code{confint}
applied to a \code{glmer} fit will first fit the profile,
then use it to compute profile confidence intervals.  Given
the computational cost of profiling, it makes sense to compute
and save the profile as an intermediate step if one plans to
do anything other than computing confidence intervals.)

Two other common methods for computing confidence intervals
are parametric bootstrapping (\code{method = "boot"}) and the
classical Wald approximation (\code{method = "Wald"}).
Parametric bootstrapping is much slower, but more accurate
(and, via the \code{FUN} argument, can generate confidence intervals
for any quantity that can be derived from a fitted model).
The Wald approximation is faster and less accurate than profile
confidence intervals.
By default \code{glmer} only returns estimates for the fixed-effects
parameters, as the assumptions of the Wald approximation are often
violated badly for random-effects (co)variances and correlations.
In the examples below we use the finite-difference Hessian
(second derivative matrix of the estimated parameters) and the
delta method to compute Wald confidence intervals for random-effects
standard deviations and correlations, when possible.

Figure~\ref{fig:cbppcompplot} compares all three of these confidence intervals
across all three of the models fitted.

\begin{knitrout}
\definecolor{shadecolor}{rgb}{0.969, 0.969, 0.969}\color{fgcolor}\begin{figure}
\includegraphics[width=\maxwidth]{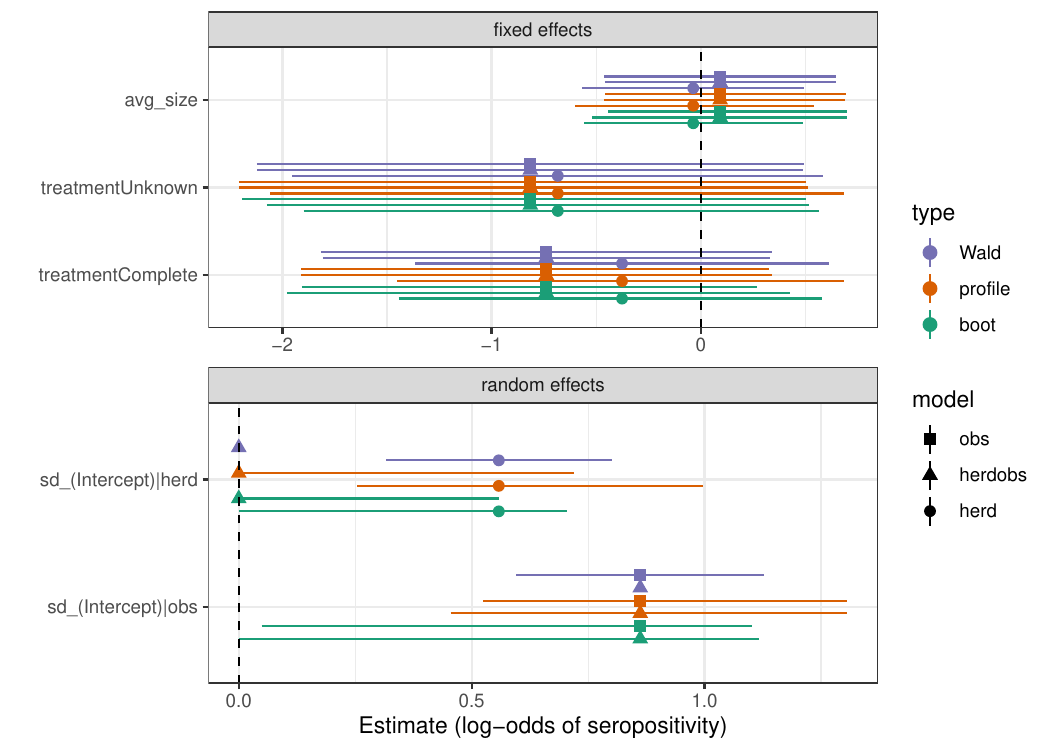} \caption[CBPP comparison]{CBPP example: comparison of point and confidence interval estimation for different methods. Wald CIs are missing for random-effects parameters when the fitted model is singular.}\label{fig:cbppcompplot}
\end{figure}

\end{knitrout}

If the default optimizers (BOBYQA followed by Nelder-Mead) do not
perform well, one could attempt to re-fit the model with a variety of 
different optimizers using \code{allFit()}.
The command \code{allFit(show.meth.tab=TRUE)} lists the optimizers \code{lme4}
currently supports.

To use \code{allFit()}, supply the initial fitted model as input. Useful
results are obtained from \code{summary(allFit())}; the components below
(other than \code{\$which.OK}) apply only to the optimizers that succeeded:
\begin{itemize}
  \item \code{\$which.OK} --- which optimizers worked
  \item \code{\$llik} --- log-likelihoods
  \item \code{\$fixef} --- fixed-effect estimates
  \item \code{\$sdcor} --- random-effect standard deviations and correlations
  \item \code{\$theta} --- random-effect parameters on the Cholesky scale
\end{itemize}

We will show the summary results for \code{\$sdcor}; the other components are similar.

\begin{knitrout}
\definecolor{shadecolor}{rgb}{0.969, 0.969, 0.969}\color{fgcolor}\begin{kframe}
\begin{alltt}
\hldef{> }\hldef{gm_all} \hlkwb{<-} \hlkwd{allFit}\hldef{(gm1)}
\end{alltt}
\begin{verbatim}
bobyqa : [OK]
Nelder_Mead : [OK]
nlminbwrap : [OK]
nmkbw : 
[OK]
optimx.L-BFGS-B : [OK]
nloptwrap.NLOPT_LN_NELDERMEAD : [OK]
nloptwrap.NLOPT_LN_BOBYQA : [OK]
\end{verbatim}
\end{kframe}
\end{knitrout}

\begin{knitrout}
\definecolor{shadecolor}{rgb}{0.969, 0.969, 0.969}\color{fgcolor}\begin{kframe}
\begin{alltt}
\hldef{> }\hldef{ss} \hlkwb{<-} \hlkwd{summary}\hldef{(gm_all)}
\hldef{> }\hldef{ss}\hlopt{$}\hldef{sdcor}
\end{alltt}
\begin{verbatim}
                              herd.(Intercept)
bobyqa                               0.5581743
Nelder_Mead                          0.5581777
nlminbwrap                           0.5581803
nmkbw                                0.5580124
optimx.L-BFGS-B                      0.5581753
nloptwrap.NLOPT_LN_NELDERMEAD        0.5581391
nloptwrap.NLOPT_LN_BOBYQA            0.5581637
\end{verbatim}
\end{kframe}
\end{knitrout}

As of version 2.0, \code{lme4} can also specify structured variance-covariance 
matrices for (generalized) linear mixed models.
\code{lme4} now supports unstructured (general positive definite), diagonal, 
compound symmetry, and first-order autoregressive (AR1) structures.
By default, AR1 models assume a homogeneous-variance model (the variance is the same for all time steps), while the other models assume heterogeneous-variance models (variances differ for every level of the varying term); users can adjust this with the \code{hom} argument (e.g. \code{ar1(..., hom = FALSE)}).

The unstructured covariance structure is the default for mixed models.
Here we illustrate fitting an AR1 model; the next example will show a compound
symmetric model.

\begin{knitrout}
\definecolor{shadecolor}{rgb}{0.969, 0.969, 0.969}\color{fgcolor}\begin{kframe}
\begin{alltt}
\hldef{> }\hldef{gm.ar1} \hlkwb{<-} \hlkwd{glmer}\hldef{(incidence}\hlopt{/}\hldef{size} \hlopt{~} \hlkwd{ar1}\hldef{(}\hlnum{1} \hlopt{+} \hldef{herd} \hlopt{|} \hldef{period),}
\hldef{+ }                \hlkwc{family} \hldef{= binomial,}
\hldef{+ }                \hlkwc{data} \hldef{= cbpp,} \hlkwc{weights} \hldef{= size)}
\hldef{> }\hlkwd{print}\hldef{(}\hlkwd{VarCorr}\hldef{(gm.ar1))}
\end{alltt}
\begin{verbatim}
 Groups Name        Std.Dev. Corr        
 period (Intercept) 0.78959  -0.041 (ar1)
\end{verbatim}
\end{kframe}
\end{knitrout}

For this particular model, a heterogeneous AR1 model (\code{ar1(..., hom = FALSE)}) results in a singular fit.

\subsection{Contraception}

\citet{huq1990bangladesh} use multilevel models to analyze data from a
fertility survey of women in Bangladesh. These data are available as
the \code{Contraception} object in the \pkg{mlmRev} package. The
response variable is binary and indicates whether or not each woman
was using contraception at the time of the survey. Covariates included
the woman's age, the number of live children she had, whether she
lived in an urban or rural setting, and the district in which she
lived.

\begin{knitrout}
\definecolor{shadecolor}{rgb}{0.969, 0.969, 0.969}\color{fgcolor}\begin{figure}
\includegraphics[width=\maxwidth]{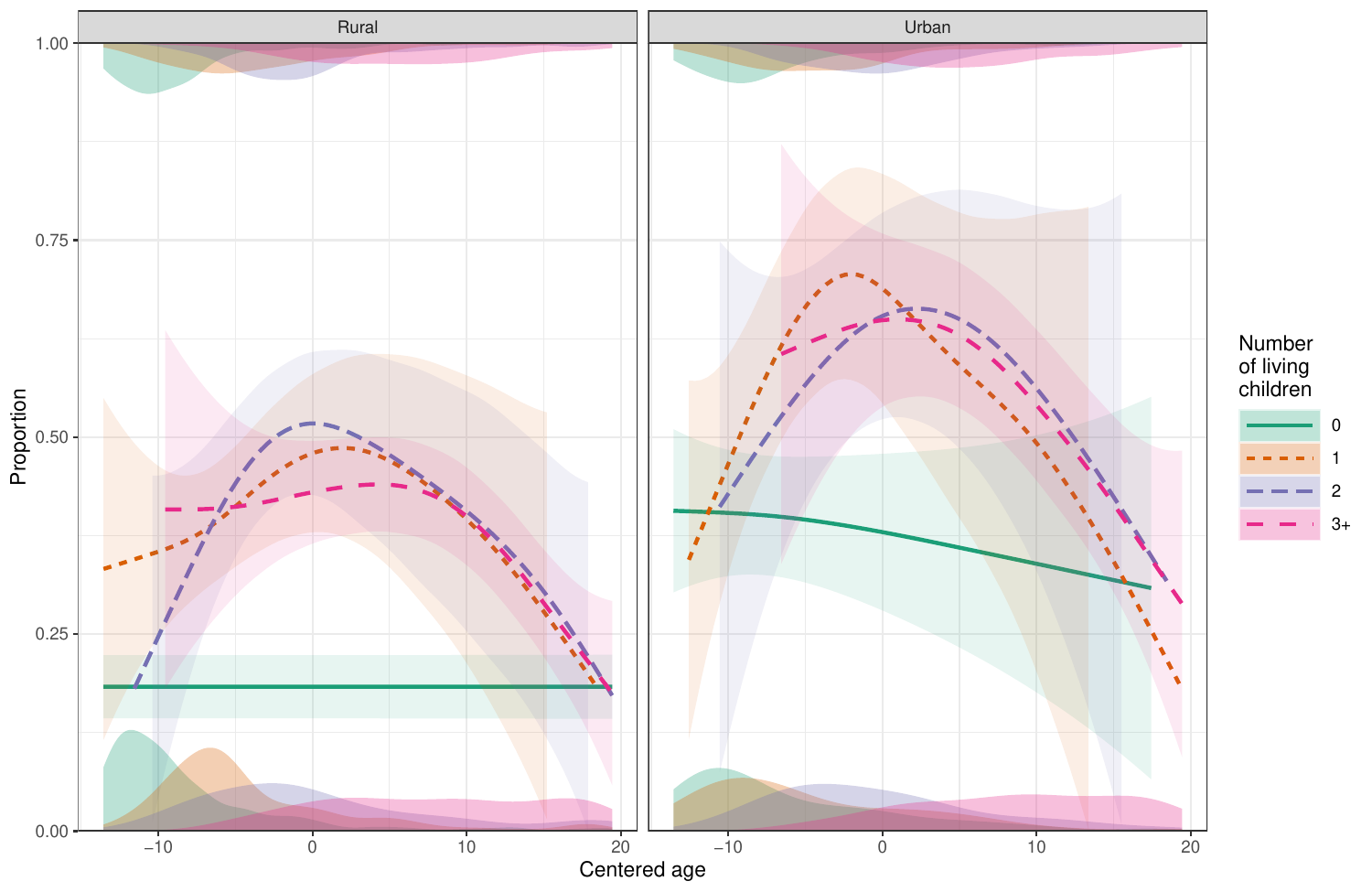} \caption[Contraception example]{Contraception example: proportion of contraceptive use by centered age, number of living children (line type), and urban/rural residence (facet). Curves fitted using generalized additive models with cubic spline smoothing. Density plots along bottom and top margins show the age distributions of women not using contraception (bottom) and using contraception (top).}\label{fig:graphContraception}
\end{figure}

\end{knitrout}

Figure~\ref{fig:graphContraception} shows exploratory smooth curves of
contraceptive use as a function of centered age, stratified by number
of living children and urban/rural residence. In rural areas, women
with two living children show the highest rates of contraceptive use,
peaking near 50\% at the average age, while women in urban areas with
one living child show the highest rates near the average
age but with a more pronounced decline at older ages. In both
settings, women with no living children consistently show the lowest
rates of contraceptive use. The nonmonotonic effect of age on contraceptive use
and the apparent dependence of this age trend on child status motivate the
model specifications explored below. (These nonmonotonic trends were not
noticed in the original analysis of the data, which concluded that there
was no significant effect of age.)

We construct six \code{glmer} models with varying choices of explanatory
variables and random effects structure, comparing them via \code{anova}. The
models considered different variables.

For instance, there are two different ways to encode the number of living 
children: \code{livch} is a four-level factor distinguishing \code{0}, \code{1}, 
\code{2}, or \code{3+} children, while other models use \code{ch} instead, which is a binary indicator for whether the 
woman has any living children (in Table~\ref{tab:contraictab} and Figure~\ref{fig:contracompplot}
we label this variable as \code{binary\_child}).

Second, we vary whether child status interacts with the woman's age: two models
include \code{ch} (or \code{livch}) and \code{age} as additive terms, while the 
other four include a \code{ch} and \code{age} interaction. A quadratic effect of
age (\code{I(age\^{}2)}) was added to account for the nonlinear effect of age.

Third, we explore different random effects structures at the district level.
Three of them use a single random intercept per district \code{(1 | district)}.
One of them extends this with a random slope for urban status 
\code{(urban | district)}, allowing the urban/rural difference to vary by district. 
The next uses nested random intercepts for district and site within district (\code{1 | district/urban})
separating district-level from urban-within-district variation.
(\cite{scandola2024} refer to this formulation as a ``complex random intercepts'' model).
The remaining model uses only the \code{urban:district} grouping. 
To reduce convergence warnings and facilitate interpretability, age (already approximately centered
in the data set) was standardized by scaling by twice the standard deviation \citep{gelman2008scaling}.

\begin{knitrout}
\definecolor{shadecolor}{rgb}{0.969, 0.969, 0.969}\color{fgcolor}\begin{kframe}
\begin{alltt}
\hldef{> }\hldef{cm1} \hlkwb{<-} \hlkwd{glmer}\hldef{(use} \hlopt{~} \hldef{age_s} \hlopt{+} \hlkwd{I}\hldef{(age_s}\hlopt{^}\hlnum{2}\hldef{)} \hlopt{+} \hldef{urban} \hlopt{+} \hldef{livch} \hlopt{+} \hldef{(}\hlnum{1}\hlopt{|}\hldef{district),}
\hldef{+ }                           \hldef{Contraception, binomial)}
\hldef{> }\hlcom{## switch from livch (ordinal) to ch (binary)}
\hldef{> }\hldef{cm2} \hlkwb{<-} \hlkwd{update}\hldef{(cm1, .} \hlopt{~} \hldef{.} \hlopt{-} \hldef{livch} \hlopt{+} \hldef{ch)}
\hldef{> }\hlcom{## add age by children interaction}
\hldef{> }\hldef{cm3} \hlkwb{<-} \hlkwd{update}\hldef{(cm2, .} \hlopt{~} \hldef{.} \hlopt{+} \hldef{age_s}\hlopt{:}\hldef{ch)}
\hldef{> }\hlcom{## allow urban effect to vary across districts (correlated)}
\hldef{> }\hldef{cm4} \hlkwb{<-} \hlkwd{update}\hldef{(cm3, .} \hlopt{~} \hldef{.} \hlopt{-} \hldef{(}\hlnum{1}\hlopt{|}\hldef{district)} \hlopt{+} \hldef{(}\hlnum{1}\hlopt{+}\hldef{urban}\hlopt{|}\hldef{district))}
\hldef{> }\hlcom{## compound symmetric/nested formulation}
\hldef{> }\hldef{cm5} \hlkwb{<-} \hlkwd{update}\hldef{(cm3, .} \hlopt{~} \hldef{.} \hlopt{-} \hldef{(}\hlnum{1}\hlopt{|}\hldef{district)} \hlopt{+} \hldef{(}\hlnum{1} \hlopt{|} \hldef{district}\hlopt{/}\hldef{urban))}
\hldef{> }\hlcom{## as above but drop district effect}
\hldef{> }\hldef{cm6} \hlkwb{<-} \hlkwd{update}\hldef{(cm3, .} \hlopt{~} \hldef{.} \hlopt{-} \hldef{(}\hlnum{1}\hlopt{|}\hldef{district)} \hlopt{+} \hldef{(}\hlnum{1} \hlopt{|} \hldef{district}\hlopt{:}\hldef{urban))}
\end{alltt}
\end{kframe}
\end{knitrout}

\begin{table}

\begin{tabular}{l|r|r|r}
\hline
  & df & Δnegloglik & ΔAIC\\
\hline
binary\_child × age + (1 | district:urban) & 7 & 0.472 & 0.00\\
\hline
binary\_child × age + (1 | district/urban) & 8 & 0.467 & 1.99\\
\hline
binary\_child × age + (1 + urban | district) & 9 & 0.000 & 3.06\\
\hline
binary\_child × age + (1 | district) & 7 & 5.825 & 10.71\\
\hline
binary\_child + age  + (1 | district) & 6 & 9.828 & 16.71\\
\hline
int\_child    + age  + (1 | district) & 8 & 9.599 & 20.25\\
\hline
\end{tabular}

\caption{Model comparison for Contraception fits. Note that for likelihood ratio tests, or for AIC comparisons restricted to nested models \citep{ripleySelectingAmongstLarge2004a}, the nesting sequence for the random-effects models is \code{(1+urban|district)} $>$ \code{(1|district/urban)} $>$ \{%
\code{(1|district)}, \code{(1|district:urban)}%
\}.
}
\label{tab:contraictab}
\end{table}

Table~\ref{tab:contraictab} shows that
the top three models, all of which include a child-by-age interaction and some effect of urbanization, fit approximately equally well ($\Delta$ negative log-likelihood $<0.5$). The \code{(1|district/urban)} model barely improves on the fit of \code{(1|district:urban)} (0.005 log-likelihood units), at the cost of an extra variance parameter, so it is almost 2 AIC units worse. \code{(1 + urban | district)} is a bit better ($\approx$ 0.5 log-likelihood units), but includes a covariance parameter.
Models that drop the interaction between child status and age or replace the binary child indicator with an integer count perform substantially worse ($\Delta \textrm{AIC} > 10$).

Overall, the results suggest that accounting for urban/rural variation at the district level and including the age and child interaction are both important, but the precise random effects structure for urban/rural variation is relatively
unimportant.

\begin{knitrout}
\definecolor{shadecolor}{rgb}{0.969, 0.969, 0.969}\color{fgcolor}\begin{figure}
\includegraphics[width=\maxwidth]{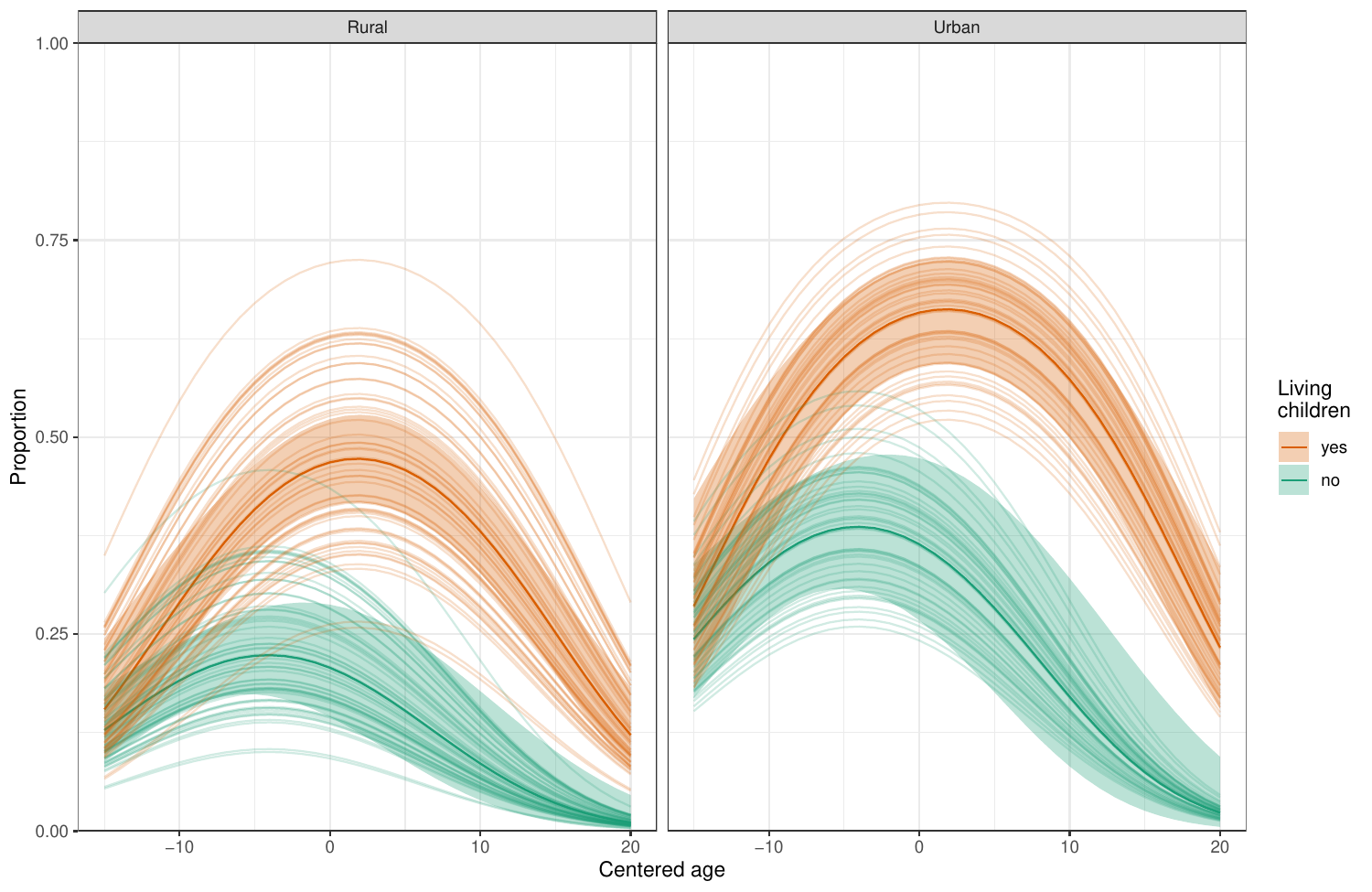} \caption[Predictions for contraception fit (\code{(1|district/urban)} model)]{Predictions for contraception fit (\code{(1|district/urban)} model). Heavy lines and ribbons show population-level predictions; light lines show district-level predictions.}\label{fig:contr-predplot}
\end{figure}

\end{knitrout}

As with the CBPP data set, we can also compute
profile, Wald, and parametric bootstrap confidence intervals for all of the models to
understand the effects of each variable and visualize the among-model variation.

\begin{knitrout}
\definecolor{shadecolor}{rgb}{0.969, 0.969, 0.969}\color{fgcolor}\begin{figure}
\includegraphics[width=\maxwidth]{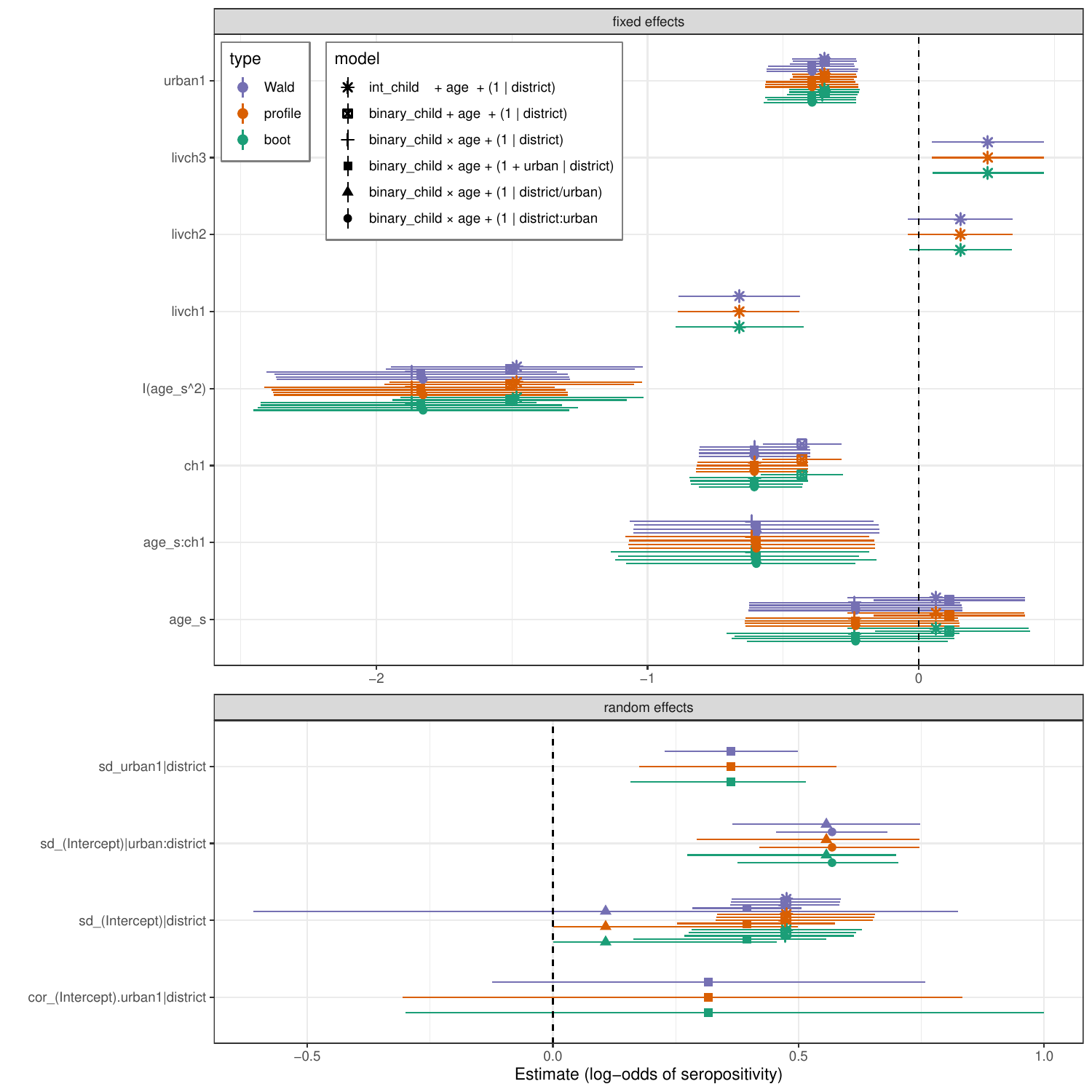} \caption[Contraception comparison]{Contraception example: comparison of point and confidence interval estimation for different methods. (Note that the Wald CI for variation in the intercept across districts, in the \code{(1|district/urban)} model, includes negative values.)}\label{fig:contracompplot}
\end{figure}

\end{knitrout}

The point estimates and confidence intervals of the explanatory variables are similar across the six models, and across different methods for confidence interval construction, with a few exceptions. 

Urban residence, presence of living children, and age were all strong predictors of contraceptive use among women in Bangladesh (because we are fitting a quadratic model for the effect of age, the non-significant effect of \code{age\_s} simply means that the marginal effect of age \emph{at the mean age} is not clearly negative or positive).

Returning to the structured covariance matrices introduced in
Section~\ref{sec:cbpp}, we can replace the unstructured random effect
\code{(1+urban|district)} with diagonal (\code{diag}) or compound symmetric
(\code{cs}) structures, with either heterogeneous or homogeneous variances
(via the \code{hom} argument):

For example, the compound symmetric model is fitted as follows:

\begin{knitrout}
\definecolor{shadecolor}{rgb}{0.969, 0.969, 0.969}\color{fgcolor}\begin{kframe}
\begin{alltt}
\hldef{> }\hldef{cm.cs} \hlkwb{<-} \hlkwd{glmer}\hldef{(use} \hlopt{~} \hldef{age}\hlopt{*}\hldef{ch} \hlopt{+} \hlkwd{I}\hldef{(age}\hlopt{^}\hlnum{2}\hldef{)} \hlopt{+} \hldef{urban} \hlopt{+} \hlkwd{cs}\hldef{(}\hlnum{1} \hlopt{+} \hldef{urban} \hlopt{|} \hldef{district),}
\hldef{+ }               \hldef{Contraception, binomial,} \hlkwc{control} \hldef{= cc)}
\end{alltt}
\end{kframe}
\end{knitrout}

We can use \code{VarCorr} and other accessor methods as we would for models with default, unstructured covariance matrices, e.g.:

\begin{knitrout}
\definecolor{shadecolor}{rgb}{0.969, 0.969, 0.969}\color{fgcolor}\begin{kframe}
\begin{alltt}
\hldef{> }\hlkwd{print}\hldef{(}\hlkwd{VarCorr}\hldef{(cm.cs))}
\end{alltt}
\begin{verbatim}
 Groups   Name        Std.Dev. Corr      
 district (Intercept) 0.615    -0.79 (cs)
          urbanY      0.725              
\end{verbatim}
\end{kframe}
\end{knitrout}

\section*{Acknowledgements}

We would like to thank Steve Walker, for early work on \code{glmer}, and Alex Stringer, for helpful comments on the manuscript.

\bibliography{glmer}

\section*{Package versions used}

Compiled with R Under development (unstable) (2026-06-29 r90199) and package versions lme4: 2.0.4, performance: 0.17.1, DHARMa: 0.5.0, see: 0.14.1.

\section{Appendix: derivation of PIRLS}

We seek to maximize the unscaled conditional log density for a GLMM
over the conditional modes, $\bm u$. This problem is very similar to
maximizing the log-likelihood for a GLM, which is a very
thoroughly studied problem \citep[e.g.][]{McCullaghNelder1989}.
The standard algorithm for dealing with this kind of
problem is iteratively reweighted least squares (IRLS). Here we modify
IRLS by incorporating a penalty term that accounts for variation in
the random effects; we call the resulting algorithm
penalized iteratively reweighted
least squares (PIRLS).

The unscaled conditional log-density takes the form,
\begin{equation}
  f(\bm u) = \log p(\bm y, \bm u | \bm\beta, \bm\theta) = 
  \bm\psi^\top \bm A \bm y - 
  \bm a^\top \bm \phi  + 
  \bm c -
  \frac{1}{2}\bm u^\top \bm u -
  \frac{q}{2}\log{2\pi}
  \label{eq:unsccondlogdens}
\end{equation}
where $\bm\psi$ is the $n$-by-$1$ canonical parameter of an
exponential family, $\bm\phi$ is the $n$-by-$1$ vector of cumulant
functions, $\bm c$ an $n$-by-$1$ vector of normalizing constants, and
$\bm A$ is an $n$-by-$n$ diagonal matrix of prior weights, $\bm
a$. Both $\bm a$ and $\bm c$ could depend on a dispersion parameter,
although we ignore this possibility for now.

The canonical parameter, $\bm\psi$, and vector of cumulant functions,
$\bm\phi$, depend on a linear predictor,
\begin{equation}
  \bm\eta = \bm o + \bm X \bm\beta + \bm Z \bm\Lambda_\theta \bm u
\end{equation}
where $\bm o$ is an $n$-by-$1$ vector of \emph{a priori} offsets. The
specific form of this dependency is specified by the choice of the
exponential family. The mean of this distribution, $\bm\mu$,
is the \emph{inverse link function} $g^{-1}$ applied to $\bm\eta$.

Our goal is to find the values of $\bm u$ that maximize the unscaled
conditional density, for given $\bm\theta$ and $\bm\beta$
vectors. These maximizers are the conditional modes, which we require
for the Laplace approximation and adaptive Gauss-Hermite
quadrature. To do this maximization we use a variant of the Fisher
scoring method, which is the basis of the iteratively reweighted least
squares algorithm for generalized linear models. Fisher scoring is
itself based on Newton's method, which we apply first.

\subsection{Newton's method}

To apply Newton's method, we need the gradient and the Hessian of the
unscaled conditional log-likelihood. Following standard GLM theory
\citep{McCullaghNelder1989}, we use the chain rule,
\begin{displaymath}
  \frac{d L(\bm\beta, \bm\theta | \bm y, \bm u)}{d \bm u} = 
  \frac{d L'(\bm\beta, \bm\theta | \bm y, \bm u)}{d \bm\psi}
  \frac{d \bm\psi}{d \bm\mu}
  \frac{d \bm\mu}{d \bm\eta}
  \frac{d \bm\eta}{d \bm u} - \frac{1}{2} \frac{d \, (\bm u^\top \bm u)}{d \, \bm u}
\end{displaymath}
where $L'$ represents the log-density in (\ref{eq:unsccondlogdens}) exclusive of the penalty term.
The first derivative in the first term's chain follows from basic results in GLM
theory,
\begin{displaymath}
  \frac{d L'(\bm\beta, \bm\theta | \bm y, \bm u)}{d \bm\psi} = 
  (\bm y - \bm\mu)^\top \bm A .
\end{displaymath}
Again from standard GLM theory, the next two derivatives define the
inverse diagonal variance matrix,
\begin{displaymath}
  \frac{d \bm\psi}{d \bm\mu} = \bm V^{-1}
\end{displaymath}
and the diagonal Jacobian matrix,
\begin{displaymath}
\frac{d \bm\mu}{d \bm\eta} = \bm M \quad .
\end{displaymath}
Finally, because $\bm u$ affects $\bm\eta$ only linearly,
\begin{displaymath}
  \frac{d \bm\eta}{d \bm u} = \bm Z \bm\Lambda_\theta
\end{displaymath}
Therefore we have,
\begin{equation}
  \frac{d L(\bm\beta, \bm\theta | \bm y, \bm u)}{d \bm u} = 
  (\bm y - \bm\mu)^\top \bm A
  \bm V^{-1}
  \bm M
  \bm Z \bm\Lambda_\theta -
  \bm u^\top \quad .
\label{eq:dPDEVdu}
\end{equation}
This is very similar to the gradient for GLMs with respect to fixed
effects coefficients, $\bm\beta$. The only difference induced by
including the penalty term for the random effects ($\bm u$), is the
subtraction of the $\bm u^\top$ term.

Again we apply the chain rule to take the Hessian,
\begin{equation}
  \frac{d^2 L(\bm\beta, \bm\theta | \bm y, \bm u)}{d \bm u \, d \bm u} = 
  \frac{d^2 L'(\bm\beta, \bm\theta | \bm y, \bm u)}{d \bm u \, d \bm\mu}
  \frac{d \bm\mu}{d \bm\eta}
  \frac{d \bm\eta}{d \bm u} - \bm I_q
\end{equation}
which leads to,
\begin{equation}
  \frac{d^2 L(\bm\beta, \bm\theta | \bm y, \bm u)}{d \bm u \, d \bm u} = 
  \frac{d^2 L'(\bm\beta, \bm\theta | \bm y, \bm u)}{d \bm u \, d \bm\mu}\bm
  M \bm Z \bm\Lambda_\theta 
  - \bm I_q
\end{equation}
The first derivative in this chain can be expressed as,
\begin{equation}
  \frac{d^2 L'(\bm\beta, \bm\theta | \bm y, \bm u)}{d \bm u \, d \bm\mu} =
  -\bm\Lambda_\theta^\top \bm Z^\top \bm M \bm V^{-1} \bm A  + 
  \bm\Lambda_\theta^\top \bm Z^\top \left[ \frac{d \bm M \bm V^{-1}}{d \bm\mu} \right] \bm A \bm R
\end{equation}
where $\bm R$ is a diagonal residuals matrix with $\bm y-\bm\mu$ on
the diagonal. The two terms arise from a type of product rule, where
we first differentiate the residuals, $\bm y-\bm\mu$, and then the
diagonal matrix, $\bm M \bm V^{-1}$, with respect to $\bm\mu$.

The Hessian can therefore be expressed as,
\begin{equation}
  \frac{d^2 L(\bm\beta, \bm\theta | \bm y, \bm u)}{d \bm u d \bm u} = 
  -\bm \Lambda_\theta^\top \bm Z^\top \bm M \bm A^{1/2}\bm V^{-1/2}\left( 
    \bm I_n - 
    \bm V \bm M^{-1}\left[ \frac{d \bm M \bm V^{-1}}{d \bm\mu} \right] \bm R
  \right) \bm V^{-1/2}\bm A^{1/2} \bm M \bm Z \bm\Lambda_\theta - \bm I_q
\label{eq:betaHessian}
\end{equation}
This result can be simplified by expressing it in terms of a weighted
random-effects design matrix, $\bm U = \bm A^{1/2}\bm V^{-1/2}\bm M
\bm Z \bm\Lambda_\theta$,
\begin{equation}
  \frac{d^2 L(\bm\beta, \bm\theta | \bm y, \bm u)}{d \bm u \, d \bm u} = 
  -\bm U^\top\left( 
    \bm I_n - 
    \bm V \bm M^{-1}\left[ \frac{d \bm V^{-1}\bm M}{d \bm\mu} \right] \bm R
  \right) \bm U - \bm I_q \quad .
\label{eq:betaHessiansimp}
\end{equation}

\subsection{Fisher-like scoring}

There are two ways to further simplify this expression for $\bm U^\top
\bm U$. The first is to use the canonical link function for the family
being used. Canonical links have the property that $\bm V = \bm M$,
which means that for canonical links,
\begin{equation}
  \frac{d^2 L(\bm\beta, \bm\theta | \bm y, \bm u)}{d \bm u \, d \bm u} = 
  -\bm U^\top\left( 
    \bm I_n - 
    \bm I_n \left[ \frac{d \bm I_n}{d \bm\mu} \right] \bm R
  \right) \bm U - \bm I_q = - \bm U^\top \bm U - \bm I_q
\end{equation}
The second way to simplify the Hessian is to take its expectation with
respect to the distribution of the response, conditional on the
current values of the spherical random effects coefficients, $\bm
u$. The diagonal residual matrix, $\bm R$, has expectation
0. Therefore, because the response only enters into the expression for
the Hessian via $\bm R$, we have that,
\begin{equation}
  E\left(\frac{d^2 L(\bm\beta, \bm\theta | \bm y, \bm u)}{d \bm u \, d \bm
      u} | \bm u \right) = 
  -\bm U^\top\left( 
    \bm I_n - 
    \bm U \bm M^{-1}\left[ \frac{d \bm V^{-1}\bm M}{d\mu} \right] E(\bm R)
  \right) \bm U - \bm I_q = - \bm U^\top \bm U - \bm I_q
\label{eq:simphesstwo}
\end{equation}

Like base R's \code{glm}, \code{glmer} uses Fisher scoring throughout (which is identical to Newton's method for canonical links) to enable the use of the (P)IRLS algorithm. As usual, we minimize the conditional negative log-density rather than maximizing the conditional log-density.

\end{document}